\documentclass[a4paper,twocolumn,superscriptaddress,11pt,accepted=2018-02-12]{quantumarticle}
\usepackage{graphicx,color}
\usepackage{amsmath,amsfonts,enumerate,amsthm,amssymb,bbm}
\usepackage{hyperref}
\usepackage{multirow}
\usepackage{bbold}
\usepackage{braket}
\usepackage{float}
\usepackage{dblfloatfix}
\usepackage[caption = false]{subfig}
\usepackage[a4paper, total={9in, 9.5in}]{geometry}
\usepackage{ulem}
\usepackage{nicefrac}
\usepackage[numbers,sort&compress]{natbib}

\pdfoutput=1
\usepackage[utf8]{inputenc}
\usepackage[english]{babel}
\usepackage[T1]{fontenc}
\usepackage{amsmath}
\usepackage{hyperref}
\usepackage{tikz}

\usepackage{scrextend}

\mathchardef\ordinarycolon\mathcode`\:
\mathcode`\:=\string"8000
\begingroup \catcode`\:=\active
  \gdef:{\mathrel{\mathop\ordinarycolon}}
\endgroup

\hypersetup{
   colorlinks=true,         
   linkcolor=blue,          
   citecolor=blue,          
}

\theoremstyle{plain}
\newtheorem{thm}{Theorem}

\theoremstyle{assumpt}
\newtheorem{assumpt}{Assumption}
\theoremstyle{definition}
\newtheorem{defn}{Definition}
\theoremstyle{remark}

\def\<{\langle}

\def\P{ {\cal P} }

\def\P{ {\cal P} }

\def\T{ {\cal T} }

\def\N{ {\cal N} }

\def\I{ \mathbb{1} }

\def\I{ \mathbbm{1} }

\def\>{\rangle}
\def\<{\langle}
\DeclareMathOperator{\Tr}{Tr}

\newcommand{\be}{\begin{equation}}
\newcommand{\ee}{\end{equation}}

\newcommand{\ja}[1]{{\color[rgb]{0,0,0}{#1}}}

\newcommand{\change}[1]{{\color[RGB]{0,0,0}{#1}}}

\newcommand{\flo}[1]{{\color[RGB]{0,0,0}{#1}}}
\newcommand{\zo}[1]{{\color[RGB]{0,0,0}{#1}}}

\newcommand{\zrewrite}[1]{{\color[RGB]{0,0,0}{#1}}}
\newcommand{\coh}[1]{{\color[RGB]{0,0,0}{#1}}}
\begin{document}

\title{Coherent fluctuation relations: from the abstract to the concrete}

\author{Zo\"e Holmes}
\affiliation{Controlled Quantum Dynamics Theory Group, Imperial College London, London, SW7 2BW, United Kingdom.}
\author{Sebastian Weidt}
\affiliation{Department of Physics and Astronomy, University of Sussex, Brighton BN1 9QH, United Kingdom.}
\author{David Jennings}
\affiliation{Controlled Quantum Dynamics Theory Group, Imperial College London, London, SW7 2BW, United Kingdom.}
\affiliation{Department of Physics, University of Oxford, Oxford, OX1 3PU, United Kingdom.}
\affiliation{School of Physics and Astronomy, University of Leeds, Leeds, LS2 9JT, United Kingdom.}
\author{Janet Anders}
\affiliation{CEMPS, Physics and Astronomy, University of Exeter, Exeter, EX4 4QL, United Kingdom.}
\author{Florian Mintert.}
\affiliation{Controlled Quantum Dynamics Theory Group, Imperial College London, London, SW7 2BW, United Kingdom.}

\begin{abstract}
Recent studies using the quantum information theoretic approach to thermodynamics show that the presence of coherence in quantum systems generates corrections to classical fluctuation theorems.
To explicate the physical origins and implications of such corrections, we here convert an abstract framework of an autonomous quantum Crooks relation into quantum Crooks equalities for well-known coherent, squeezed and cat states. We further provide a proposal for a concrete experimental scenario to test these equalities. 
Our scheme consists of the autonomous evolution of a trapped ion and uses a position dependent AC Stark shift.
\end{abstract}

\maketitle

\section{Introduction}
The emergent field of quantum thermodynamics seeks to extend the laws of thermodynamics and non-equilibrium statistical mechanics to quantum systems. A central question is whether quantum mechanical phenomena, such as coherence and entanglement, generate corrections to classical thermal physics.

A quantum information theoretic approach has proven a fruitful means of incorporating genuinely quantum mechanical effects into thermal physics~\cite{litrev}. For example, the consequences of quantum entanglement for Landauer erasure~\cite{entanglementnegentropy}, the thermodynamic arrow of time~\cite{entanglementarrowoftime} and thermalisation~\cite{entanglementthermalisation} have been investigated. In addition, for incoherent quantum systems, general criteria for state conversion~\cite{ThermodynamicTransformationsandworkextraction}, generalisations of the second laws~\cite{2ndlaws}, and limits to work extraction protocols~\cite{workextractionaberg} have been established. More recently, these results have been extended to coherent quantum states~\cite{completestateinterconversion, constraintsbeyondfreeenergy, catalyticcoherence, catalyticworkextraction}. Much of this research~\cite{ThermodynamicTransformationsandworkextraction,2ndlaws,workextractionaberg,completestateinterconversion, constraintsbeyondfreeenergy, catalyticcoherence, catalyticworkextraction} utilised the resource theory framework and in particular the concept of thermal operations~\cite{resourcetheory1,resourcetheory2}. However, despite significant theoretical progress, the results have been seen as rather abstract and unamendable to experimental implementation~\cite{manifesto}. 

A separate line of enquiry has explored fluctuation theorems, which can be seen as generalisations of the second law of thermodynamics to non-equilibrium processes~\cite{fluctreview}.
They consider systems that are driven out of equilibrium and establish exact relations between the resultant thermal fluctuations~\cite{QFT}.  Experimental tests of fluctuation theorems, in particular the Jarzynski~\cite{Jarz} and Crooks~\cite{Crooks} equalities, have been conducted in classical systems, including stretched RNA molecules~\cite{singlemolecule1,singlemolecule2}, over-damped colloidal particles in harmonic potentials~\cite{colloidalparticle} \flo{and} classical two-state systems~\cite{Pekola}, and quantum systems such as trapped ions~\cite{an} and NMR systems~\cite{nmr}. An experiment to test a quantum Jarzynski equality in a weakly measured system using circuit QED has recently been performed~\cite{Murch}.

However,  the use of two point energy measurements or continual weak measurements to obtain a work probability distribution reduces a system's
coherence with respect to the energy eigenbasis~\cite{Tasaki,Kurchan,TasakiCrooks, albash,huberjar,circuitQED}. This limits the extent to which these previous experiments probe quantum mechanical phenomena that arise from coherences.

A recent theoretical proposal of a new quantum fluctuation relation~\cite{aberg}
that connects naturally with quantum information theory, models the work system explicitly as a quantum battery. Rather than implicitly appealing to an additional classical system to drive the system out of equilibrium, the quantum system and battery here evolve \textit{autonomously} under a time independent Hamiltonian~\cite{feynmannratchet,clocks,deffjarzinclusive, autonomousfluc}. This proposal does not require projective measurements onto energy eigenstates and so does not destroy coherences. As a result coherence actively contributes to the Autonomous Quantum Crooks equality (AQC)~\cite{aberg}, which will be defined in Eq.~\eqref{eq:qcrooks}.

In this paper, we exploit the combined strengths of the quantum information and fluctuation theorem approaches to quantum thermodynamics~\cite{qifluc1,alvaro}. We develop quantum Crooks equalities for a quantum harmonic oscillator battery that is prepared in optical coherent states, squeezed states or cat states and explore the physical content of these equalities. In particular, we present the coherent state Crooks equality, see Eq.~\eqref{eq: coherentstaterewrite}, in which quantum corrections to the classical Crooks equality can be attributed to the presence of quantum vacuum fluctuations.

We propose an experiment to test the AQC through which the role of coherence in thermal physics can be probed.
Using a trapped ion, one can realise a two level system with a pair of the ion's internal energy levels and an oscillator battery with the ion's axial phonon mode. The motion of the ion through an off-resonance laser beam induces a position dependent AC Stark shift on the internal energy levels. In this way an effectively time-dependent Hamiltonian for the two level system can be realised.


\section{The Classical and Quantum Crooks Equalities} \label{Sec: The Crooks Equality}

\subsection{The classical Crooks equality}

An initially thermal system, at temperature $T$, can be driven from equilibrium with a time-dependent Hamiltonian, changing from $H_S^i$ to $H_S^f$.
The classical Crooks equality~\cite{Crooks}\flo{,}
\begin{align} \label{eq:Crooks}
\frac{ \P^+(W) }{\P^-(-W)} 
=  \exp \left( -\frac{\Delta F}{k_BT}\right) \exp \left( \frac{W}{k_BT} \right)\, ,
\end{align}
quantifies the ratio of the probability $\P^+(W)$ of the work $W$ done on a system in such a non-equilibrium process, to the probability $\P^-(-W)$ to extract the work $W$ in the time reversed process (where the Hamiltonian is changed back from $H_S^f$ to $H_S^i$). 
\zo{This ratio is a function of the change in the system's free energy, 
\begin{equation}\label{Eq: Free energy def}
\begin{aligned}
&\Delta F := F(H^f_S) - F(H^i_S) \ \ \ \  \text{with} \\
&F(H_S) : = - k_B T \,  \zo{\ln}\left( \Tr_S \left[\exp\left(-\frac{H_S}{k_B T}\right)\right] \right) \, , 
\end{aligned}
\end{equation}
and hence measurement of $P(W)$ and $P(-W)$ for a non-equilibrium process provides a means of inferring free energy changes, an equilibrium property of a system.}
More generally, Eq.~\eqref{eq:Crooks} asserts that processes that produce work are exponentially less likely than processes that require work, irrespective of how the change from $H_S^i$ to $H_S^f$ is realised. 


\subsection{The autonomous quantum Crooks equality}

The realisation of a time-dependent Hamiltonian in the classical Crooks equality, Eq.~\eqref{eq:Crooks}, necessarily implies an interaction with a classical agent. The autonomous framework~\cite{aberg} proposed by Johan \AA berg makes this control explicit by introducing a \textit{battery}.
Specifically, the system evolves together with the battery according to the time-independent Hamiltonian,
\begin{align} \label{eq: total hamiltonian}
   H_{SB} =H_S\otimes\I_B+\I_S\otimes H_B+V_{SB} \, , 
\end{align} 
comprised of the Hamiltonians $H_S$ and $H_B$ for system and battery, and their interaction $V_{SB}$.

\zo{For the initial and final Hamiltonians of the system to be well defined, we need the system and battery not to interact at these times. To ensure this, we} 
consider an interaction of the form 
\be
V_{SB} = H_S^i\otimes \Pi_B^i +  H_S^f\otimes \Pi_B^f + V_{SB}^\perp\, ,
\label{eq:interaction}
\ee
where $\Pi_B^i$ and $\Pi_B^f$
are projectors onto two orthogonal subspaces, $R_i$ and $R_f$, of the battery's Hilbert space, and $V_{SB}^\perp$ has support only outside those two subspaces, {\it i.e.} $(X_S\otimes\Pi_B^i)V_{SB}^\perp=(X_S\otimes\Pi_B^f)V_{SB}^\perp=0$ for any system operator $X_S$. The system Hamiltonian $H_S$ can always be absorbed into \zo{$V_{SB}$}, \flo{which we will do in the following.}
Assuming the battery is initialised in a state in subspace $R_i$ only and evolves to a final state in subspace $R_f$ only, the system Hamiltonian evolves from $H_S^i$ to $H_S^f$, i.e. the system Hamiltonian is effectively time-dependent.
\change{Since the system and battery are non-interacting in the regions $R_i$ and $R_f$, and since energy is globally conserved, the energy required or produced by the system as its Hamiltonian evolves from $H_S^i$ to $H_S^f$ is necessarily provided or absorbed by the battery.} 

\coh{The initially thermal system in the Crooks equality is incoherent in energy and so, to extend the relation to the quantum regime \change{in this autonomous setting}, coherence\footnote{\change{We use the term `coherence' in the sense of a `superposition of states belonging to different energy eigenspaces'. Such energetic coherences are formally quantified in the thermal operations framework \cite{resourcetheory1,resourcetheory2} and by measures such as the $l_1$ norm of coherence~\cite{coherencemeasure},  $C_{l_1}(\rho) := \sum_{j \neq k} |\bra{E_j} \rho \ket{E_k}|$, the sum of the absolute value of the off diagonal elements of a state in the energy eigenbasis.}} must begin in an additional non-equilibrium quantum system. Here the coherence is provided by the battery which we assume can be prepared in superpositions of energy states.
The evolution under $H_{SB}$ generates correlations between the system and battery reducing the coherence that can be attributed to the battery alone.
Since the dynamics conserves coherence globally,
the coherence of the final joint state of the system and battery is equal to that initially provided by the battery.
Thus this framework allows us to describe the evolution of coherence under interactions with a thermal system.}




The difficulties surrounding how to define work in the quantum regime~\cite{defofquantumwork1,defofquantumwork2,defofquantumwork3,defofquantumwork3.5,defofquantumwork4,defofquantumwork5} are avoided by formulating the AQC in terms of transition probabilities between battery states. \zo{Considering a system initialised in the thermal state $\gamma_{i}$ and the battery initialised in the pure state $\ket{\phi_i}$},
the probability to observe the battery after a time-interval $t$ in the state $\ket{\phi_f}$ reads
\be\label{eq:transitionprobabilities}
\mathcal{P}(\phi_f|\phi_i, \gamma_i)= \bra{\phi_f}\Tr_S[ U(\gamma_{i} \otimes \ket{\phi_i}\bra{\phi_i})U^\dagger]\ket{\phi_f}\, ,
\ee
with $U=\exp(-iH_{SB}t/\zo{\hbar})$ the propagator of the system and battery and the trace over the system denoted by $\Tr_S$. \coh{If we consider the transition probability between energy eigenstates, the scheme considered here reduces to the usual two-point projective energy measurement scheme~\cite{Tasaki,Kurchan,TasakiCrooks} and as such we can recover the classical Crooks relation, Eq.~\eqref{eq:Crooks}. \change{For transition probabilities between more general quantum states, such as between coherent states, squeezed states and cat states~\cite{trappedionsreviewLeibfried} as we will consider, the post measurement battery state has coherences in the energy eigenbasis and the measurement can induce a non-classical updating of the state of the system.}}

The AQC relates the probabilities $\mathcal{P}(\phi_f|\phi_i, \gamma_i)$ of a {\it forward process} and $\mathcal{P}(\psi_i|\psi_f, \gamma_f)$ of a {\it reverse  process} for two thermal \zo{\textit{system}} states
\be\label{eq:thermalstates}
\gamma_{x}  \propto \exp\left(- \frac{ H_S^x}{k_B T}\right) \, , \mbox{ with }\ x = i,f\ 
\ee
and two pairs of pure initial and final \zo{\textit{battery}} states satisfying
\be
\begin{split}
\ket{\psi_f}&\propto \T \exp\left(-\frac{H_B}{2 k_B T}\right)\ket{\phi_f}\  \, , \\
\ket{\phi_i}&\propto \T \exp\left(-\frac{H_B}{2 k_B T}\right)\ket{\psi_i}\ .
\end{split}
\label{eq:forwardtoreverse}
\ee
The states $\ket{\psi_i}$ and $\ket{\psi_f}$ correspond to a time-reversed process \ja{and} are defined in terms of the time-reversal operator\footnote{\zo{The time-reversal~\cite{timereversalop} operation $\T$ on a battery state $\ket{\psi}$ is defined as complex conjugation in the energy eigenbasis of the battery, $\T \ket{\psi} = \ket{\psi^*}$. We introduce $\T$ in more detail in Appendix~\ref{Sec: Time Reversal}.}} $\T$~\cite{timereversalop}.
\change{This relationship between the initial and final states of the forwards and reversed processes, Eq.~\eqref{eq:forwardtoreverse}, is forced by the derivation of the AQC, \textit{i.e.} the structure arises naturally when one reverses a general quantum process in an autonomous setting. If the battery states $\ket{\phi_i}$ and $\ket{\phi_f}$ are eigenstates of $H_B$, the states $\ket{\psi_f}$ and $\ket{\psi_i}$ of the reversed process are regular time-reversed states. However, for any coherent superposition of energy eigenstates the additional term $\exp(- H_B/2 k_B T)$, which is linked with the Petz recovery map~\cite{petz,hyukjoon}, is essential to capture the influence of quantum coherence.}

The AQC~\cite{aberg} reads
\be
\frac{\mathcal{P}(\phi_f|\phi_i, \gamma_i)}{\mathcal{P}(\psi_i|\psi_f, \gamma_f)}
= \exp\left(-\frac{\Delta F }{k_BT}\right)\exp\left(\frac{\Delta \tilde{E}}{k_BT}\right) \, ,
\label{eq:qcrooks}
\ee
\zo{where $\Delta F$ is the change in the system's equilibrium free energy, Eq.~\eqref{Eq: Free energy def}, and thus the AQC, similarly to the classical Crooks equality, provides a means of inferring free energy changes.} 
The term
    \be
    \Delta \tilde{E}(\psi_i, \phi_f) :=  \tilde{E}(\zo{\psi_i}) -\tilde{E}(\zo{\phi_f})
    \label{eq:flow}
    \ee
is a quantum mechanical generalisation of the energy flow from the battery to the system. The function
\be\label{Eq: tildeE}
 \tilde{E}(\zo{\psi}) := -k_B T \, \zo{\ln}\left( \zo{\bra{\psi}} \exp\left(- \frac{ H_B}{k_B T}\right) \zo{\ket{\psi}} \right)\ 
\ee
is an effective potential for the battery state $\ket{\psi}$ that specifies the relevant energy value within the fluctuation theorem context. 
We provide a discussion of its properties in Appendix~\ref{Appendix: properties of tilde}.


For the AQC to hold exactly we require $\ket{\psi_i}$ and $\ket{\phi_f}$ to have support \textit{only} within subregions $\Pi^i_B$ and $\Pi^f_B$ respectively \textit{and} for the battery Hamiltonian not to induce evolution between subregions, i.e. $ (\I_B - \Pi_B^i ) H_B \Pi_B^i = 0$ and $ (\I_B - \Pi_B^f ) H_B \Pi_B^f = 0$. In general, the second of these conditions does not hold because a battery Hamiltonian that evolves states between subregions is required. However, despite this, numerical simulations (discussed in Section \ref{Sec: Results}) indicate that any error can be made negligible as long as $\ket{\psi_i}$ and $\ket{\phi_f}$ are well localised in subregions $\Pi_B^i$ and $\Pi_B^f$ respectively. In Section \ref{sec: error bound} we discuss how the accuracy of the equality can be quantified; however, the material is relatively technical and can be skipped.
\medskip 

To make the predictions of the AQC, Eq.~\eqref{eq:qcrooks}, more concrete we now choose the battery to be a quantum harmonic oscillator with Hamiltonian $H_B = \sum_n \hbar \omega \, \left(n +\frac{1}{2}\right) \, \ket{n}\bra{n}$. 
In order to derive explicit expressions for the ratio of transition probabilities, Eq.~\eqref{eq:qcrooks}, we consider common states of the harmonic oscillator, specifically coherent states, squeezed states and cat states~\cite{trappedionsreviewLeibfried}. For each of these three states we derive a specific AQC which is
based on the generalised energy flow, Eq.~\eqref{eq:flow}, and the relation between the initial and final states of the forward and reversed process, Eq.~\eqref{eq:forwardtoreverse}. \zo{While the battery Hamiltonian and states are fixed in the following sections, the system Hamiltonians, $H_S^i$ and $H_S^f$, and the system and battery interaction, $V_{SB}^\perp \,$, \flo{can still be chosen at will}.} 

\clearpage

\subsection{Coherent state Crooks equality}\label{Sec: practicalCrooks}

The physical content of the AQC is neatly illustrated by considering transition probabilities between coherent states of the battery, $\ket{\alpha} =\exp(\alpha a^\dagger-\alpha^\ast a)\ket{0}$ with $a^\dagger$ and $a$ the creation and annihilation operators of the oscillator respectively, and $\alpha$ a complex number. In the forwards process, we assume the battery is initialised in $\ket{\alpha_i \exp(-\chi)}$ with $\chi = \frac{\hbar \omega}{2 k_B T}$ and we are interested in the transition probability $\mathcal{P}(\alpha_f | \alpha_i \exp(-\chi), \gamma_i)$ to the coherent state $\ket{\alpha_f}$. Following Eq.~\eqref{eq:forwardtoreverse}, the initial and final states of the reversed process are given by the coherent states $\ket{\alpha_f^\ast \exp(-\chi)} $ and $\ket{\alpha_i^\ast}$. The symbol $^\ast$ denotes complex conjugation in the Fock basis and arises from the time reversal operation on a coherent state, $ \T \ket{\alpha} =  \ket{\alpha^\ast}$.

The dimensionless parameter $\chi$ is the ratio of the magnitude of quantum fluctuations, $\frac{ \hbar \omega}{2}$, to the magnitude of thermal fluctuations, $k_B T$, and quantifies the degree to which any regime is quantum mechanical. 
\flo{In the limit in which thermal fluctuations dominate, $\chi $ tends to 0, and the prefactor $\exp(-\chi)$ tends to 1.}
Consequently, in this limit the reverse process is the exact time-reversed process; however, in general, the pairs of states considered in the forwards and reverse process differ by an amount determined by $\exp(-\chi)$.

As derived in more detail in \ref{Appendix: Coherent States}, the energy flow, Eq.~\eqref{eq:flow}, takes the explicit form
\begin{equation}
\begin{aligned}\label{Eq: Coherent state energy flow}
\Delta \tilde{E}(\alpha_f,\alpha_i)
= k_B T \left(|\alpha_i|^2 -|\alpha_f|^2)(1-\exp(-2\chi)\right) \ .
\end{aligned}
\end{equation}

In order to highlight similarities and differences to the classical situation, it is instructive to explicitly introduce the difference between the average energy cost $\Delta E_+$ and gain $\Delta E_-$ of the forward and reverse processes. We define the prefactor
\begin{equation}\label{eq: q factor}
q := \frac{\Delta \tilde{E}}{W_q}
\end{equation}
as the ratio between $\Delta \tilde{E}$ and $W_q=(\Delta E_+ - \Delta E_-)/2$. As shown in \ref{Appendix: Coherent States}, we find that for coherent states the prefactor takes the form
\begin{equation}\label{eq: q factor coherent}
q\left(\chi \right) = \frac{1}{\chi} \tanh(\chi) \ 
\end{equation}
and can be related to the average frequency, $\omega_T$, of the oscillator \zo{\textit{battery}} in a thermal state, \zo{$\gamma(H_B)\propto \exp\left(- \frac{ H_B}{k_B T}\right) $}, at temperature $T$, 
\begin{equation}\label{eq: omegaT}
\hbar \omega_T := \langle H_B \rangle_{\zo{\gamma(H_B)}} = \frac{k_B T }{q(\chi)} \ .
\end{equation}
The coherent state AQC can thus be written as
\begin{equation}
\begin{aligned}\label{eq: coherentstaterewrite}
&\frac{\mathcal{P}\left( \alpha_f \big\rvert  \alpha_i  \exp(-\chi) , \gamma_i \right)}{\mathcal{P}\left(\alpha_i^* \big\rvert   \alpha_f^* \exp(-\chi), \gamma_f \right)}  = \exp\left(-\frac{\Delta F}{k_B T}\right)\exp\left(\frac{W_q}{\hbar \omega_T}\right)\, .
\end{aligned}
\end{equation}
In this form, it is analogous to the classical Crooks equality, Eq.~\eqref{eq:Crooks},
with the difference encoded in $\omega_T$.

\begin{figure}
  \centering
{\includegraphics[width = 3.5in]{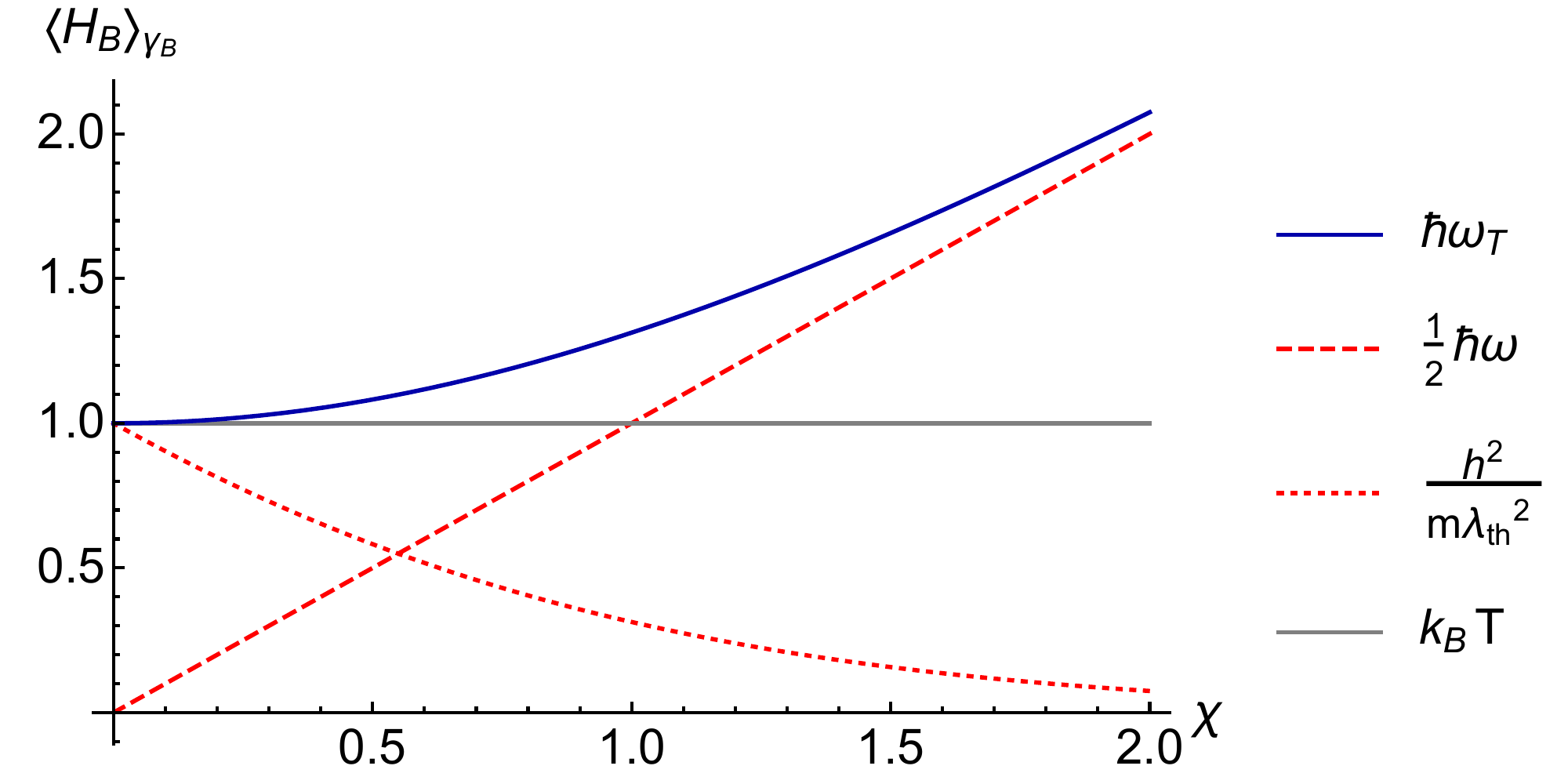}}
\caption{The solid, dark blue line shows the average energy, $\hbar \omega_T$, of the oscillator in a thermal state at temperature $T$ as a function of $\chi$. The red lines indicate the contribution of thermal (dotted) and quantum (dashed) contributions to $\hbar \omega_T$ as determined by Eq.~\eqref{eq: relating omegaT and lambdadB}. The grey line is the classical limit in which the energy of the harmonic oscillator equals $k_B T$. Energies are given in units of $k_B T$.}
\label{Fig: energy graph}
\end{figure}

The thermal frequency $\omega_{T}$ can be understood in terms of the
\flo{thermal wavelength $\lambda_{th} = \frac{h}{p}$ \cite{thermaldebrogliewavelength} given in terms of the momentum $p$ of a particle.}
This is a spatial scale that quantifies the regime in which a thermal system should be treated as classical or quantum. Broadly speaking, when $\lambda_{th}$ is small the system is well localised and can be considered classical; however, quantum effects dominate when it is larger. 

The average energy
of the oscillator is composed of kinetic, potential, and vacuum fluctuation terms.
\zo{For an oscillator in a thermal state the
kinetic and potential energies are equal, and with the vacuum energy
$E_{\mbox{\tiny  vac}} = \frac{1}{2}\hbar \omega$, 
it follows that}
\begin{equation}\label{eq: relating omegaT and lambdadB}
\hbar \omega_{T} =  \frac{h^2}{m \lambda_{th}^2} + \frac{1}{2}\hbar \omega \ .
\end{equation}
This equality describes the splitting of $\hbar \omega_T$ into its thermal, $\frac{h^2}{m \lambda_{th}^2}$, and quantum, $\frac{1}{2}\hbar \omega$, contributions, as is shown in Fig. \ref{Fig: energy graph}.


In the classical limit \flo{the thermal wavelength} $\lambda_{th}$ is small and the dominant contribution to $\hbar \omega_T$ is thermal. In this limit $\hbar \omega_T$ tends to $k_B T$ as predicted by the equipartition theorem and in agreement with the classical Crooks equality.

In the quantum limit of large $\lambda_{th}$ the thermal contribution tends to zero and vacuum fluctuations dominate. These vacuum fluctuation induce quantum corrections to $\hbar \omega_T$ which increase monotonically with $\chi$. Consequently, the exponential dependence of the ratio of transition probabilities on the difference in energy, $W_q$, is suppressed and the probability for the reverse process is larger than expected classically. We conclude that the presence of quantum coherence in a sense makes irreversibility milder. As the dynamics quantified by the AQC are unitary and so fully reversible, the irreversibility quantified here stems from the choice in the prepared and measured battery states~\cite{obversibility}.



\subsection{Squeezed state and cat state Crooks equalities}
\zo{Analogously to Eq.~\eqref{eq: coherentstaterewrite}, we now discuss quantum Crooks equalities that quantify the ratio of transition probabilities between pairs of squeezed states of the battery and pairs of cat states of the battery}. Squeezed displaced states, $\ket{r, \alpha} := \exp(\alpha a^\dagger-\alpha^\ast a) \exp(\frac{r}{2}(a^2 - {a^\dagger}^2)) \ket{0}$, are the unbalanced minimum uncertainty states where for a given squeezing parameter $r$, the variance of the oscillator's position and momentum are rescaled as $\Delta x = \exp(-r) \sqrt{\frac{\hbar}{2 m \omega}}$ and $\Delta p = \exp(+r) \sqrt{\frac{\hbar m \omega }{2}}$.
\flo{Cat states $\ket{\mbox{\small Cat}} \propto \ket{\alpha} + \ket{\beta}$ are superpositions of coherent states $\ket{\alpha}$ and $\ket{\beta}$.}

The coherent, squeezed, and cat state equalities differ in the choice of preparation and measurement states \zo{of the battery}.
The following tables list the form of these battery states respectively for the forwards ($\ket{\phi_{i/f}}$) and reverse ($\ket{\psi_{i/f}}$) processes.

\begin{center}\label{Table}
 \begin{tabular}{| c | c| c | } 
 \hline 
{\footnotesize Forwards} & {\footnotesize Preparation} & {\footnotesize Measurement }\\
  [0.5ex] 
 \hline  
{\footnotesize Coherent}  & {\footnotesize $  \ket{\alpha_i e^{-\chi}}$} & {\footnotesize $\ket{\alpha_f}$} \\ [1ex] 
{\footnotesize Squeezed} &  {\footnotesize $\ket{s_{i}, \mu_{i}}$}  & {\footnotesize  $\ket{r_{f}, \alpha_f}$ } \\ [1ex]
{ \footnotesize Cat} &{\footnotesize$ \eta_\chi^{|\alpha_i|^2} \ket{\alpha_i e^{-\chi}} + \eta_\chi^{|\beta_i|^2}\ket{ \beta_i e^{-\chi}}$}&  {\footnotesize$\ket{\alpha_f} + \ket{\beta_f}$}\\
 \hline
\end{tabular}
\end{center}

\begin{center}
 \begin{tabular}{| c | c| c | } 
 \hline
 {\footnotesize Reverse} &  {\footnotesize Preparation} & {\footnotesize Measurement} \\
  [0.5ex] 
 \hline  
{\footnotesize Coherent}  & {\footnotesize $\ket{\alpha_f^* e^{-\chi}}$} & {\footnotesize $\ket{\alpha_i^*}$} \\ [1ex] 
{\footnotesize Squeezed} &   {\footnotesize $\ket{s_{f}, \mu_{f}^*}$} &   {\footnotesize $\ket{r_{i}, \alpha_i^*}$} \\ [1ex]
{\footnotesize Cat} &  {\footnotesize $\eta_\chi^{|\alpha_f|^2} \ket{\alpha_f^* e^{-\chi} } + \eta_\chi^{|\beta_f|^2}\ket{\beta_f^* e^{-\chi}}$} & {\footnotesize $\ket{\alpha_i^*} + \ket{\beta_i^*}$} \\  
 \hline
\end{tabular}
\end{center}

In the above table the cat states are unnormalized for brevity and we have defined {\footnotesize  $\eta_\chi := \exp\left(-\frac{1}{2}\left(1- \exp(-2\chi) \right)\right)$}, {\footnotesize $s := \tanh^{-1}\left(\exp(-2\chi) \tanh(r)\right)$} and {\footnotesize $\mu := \frac{\exp(-\chi)\left(1+ \tanh(r)\right)}{1 + \exp(-2\chi) \tanh(r)}$}. We do not show the explicit expressions of $\Delta \tilde{E}$ for the cat and squeezed state equalities here as they are reasonably long and uninstructive. 
However, Appendix \ref{Appendix: Corollary Algebra} contains full derivations and statements of the two equalities. 
\medskip
\begin{figure}
\subfloat[]{\includegraphics[width = 1.65in]{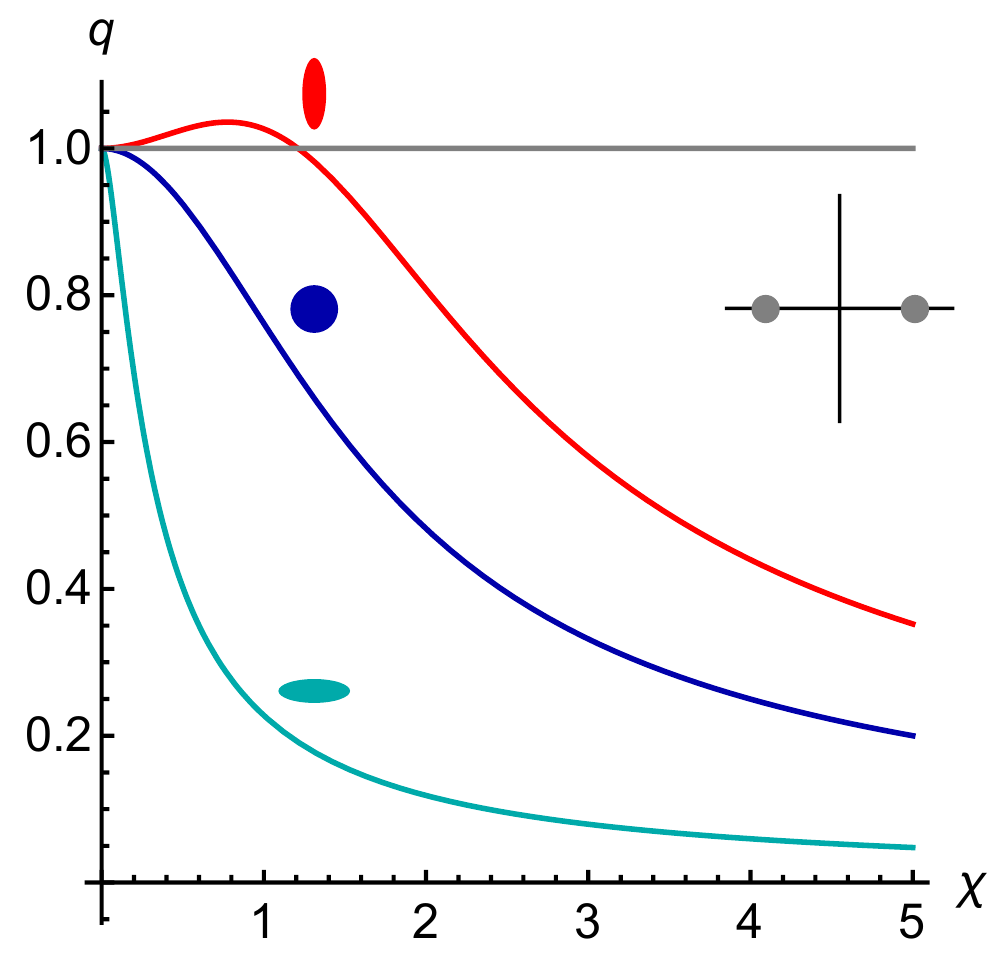}} \hspace{0.1in}
\subfloat[]{\includegraphics[width = 1.65in]{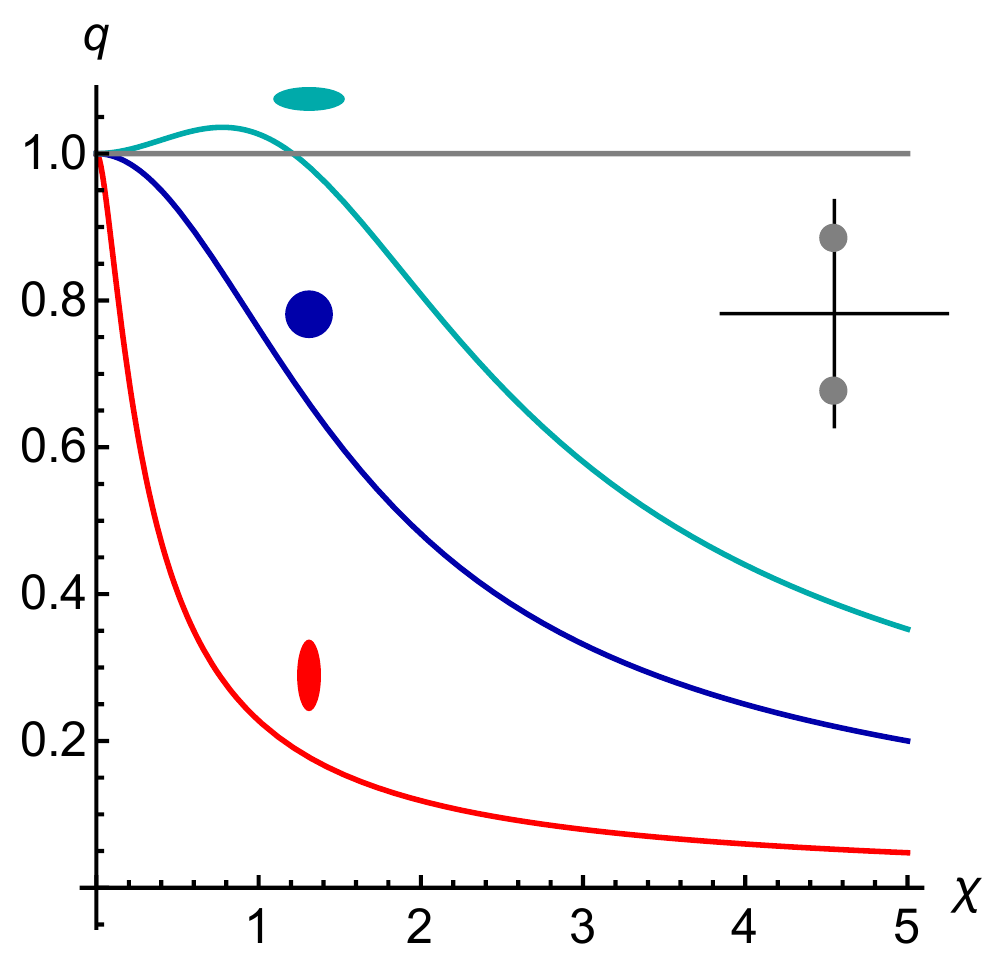}} 
\caption{We plot  $q := \frac{\Delta \tilde{E}}{W_q}$ as a function of $\chi$ for squeezed displaced states. As indicated by the grey Wigner plot insets, in a) the prepared and measured squeezed displaced battery state are displaced with respect to position only, i.e. $\alpha_{i,f} = \Re(\alpha_{i,f})$ and $\mu_{i,f} = \Re(\mu_{i,f})$; whereas in b) the state is displaced with respect to momentum only, i.e. $\alpha_{i,f} = \Im(\alpha_{i,f})$ and $\mu_{i,f} = \Im(\mu_{i,f})$. In a) the red, blue and turquoise lines plot $r = 1$, $r=0$ and $r = -1$ respectively. In b) the red, blue and turquoise lines plot $r = -1$, $r=0$ and $r = 1$ respectively. In both figures the grey line indicates the classical limit in which $q = 1$.}
\label{Fig: squeeze q factor}
\end{figure}

We compare the additional effect of squeezing to the coherent state cases in Fig. \ref{Fig: squeeze q factor} by plotting $q(\chi)$, the ratio of $\Delta \tilde{E}$ and $W_q$ as defined in Eq.~\eqref{eq: q factor}. 
As is shown in dark blue, for coherent states $q$ reduces to 1 in the classical limit where $\chi$ tends to 0.
\flo{However, $q$ decreases monotonically for increasing $\chi$, and vanishes as $\chi$ tends to infinity.}

When the battery is prepared and measured in states that are displaced with respect to position and squeezed with respect to momentum, we see in Fig. \ref{Fig: squeeze q factor}a that $q$ behaves in a similar way to the coherent state case but with a quicker initial rate of decrease.
When the battery is displaced with respect to position and squeezed with respect to position, the behaviour of $q$ for large $\chi$ again replicates the coherent state case; however, there is now a regime of low $\chi$ in which $q$ is greater than the classical value of 1. 
In Fig.~\ref{Fig: squeeze q factor}b the prepared and measured states are displaced with respect to momentum and we see identical behaviour to Fig.~\ref{Fig: squeeze q factor}a except the roles of squeezing with respect to position and momentum are reversed.

Thus we see that for a squeezed battery the probability for the reverse process is again larger than expected classically in the limit of large $\chi$. However, there is now an intermediary regime where the probability for the reverse process is even more suppressed than in the classical limit. In the presence of squeezing, irreversibility can thus be stronger than expected classically.

Cat states are inherently quantum mechanical and do not reduce to a semi-classical state in the limit of low $\chi$. Consequently, the analogous plot to Fig.~\ref{Fig: squeeze q factor} for cat states is not a straightforward correction to the classical limit. The plot is thus uninstructive and so we do not include it here.

\subsection{Quantifying accuracy of the AQC}\label{sec: error bound}

The approximate nature of the AQC can be quantified by  
error bounds. Here we establish three such error measures, $D$, $\epsilon$ and $(1-R)$. 

It is instructive to first note that Eq.~\eqref{eq:qcrooks} can be rewritten as 
\begin{equation}\label{eq: rearrangedAQC}
\begin{aligned}
&Z_i \bra{\psi_i}\exp\left(- \frac{H_B}{k_B T}\right)\ket{\psi_i} \mathcal{P}\left( \phi_f | \phi_i, \gamma_i \right) \\ &- Z_f \bra{\phi_f}\exp\left(- \frac{H_B}{k_B T}\right)\ket{\phi_f} \mathcal{P}\left(\psi_i | \psi_f, \gamma_f \right) \approx 0 \, 
\end{aligned}
\end{equation}
using the definition of the generalised energy flow, Eq.~\eqref{eq:flow}, and the definition
\flo{of the system's free energy which can be rephrased as $\exp(-\Delta F/k_B T) = Z_f / Z_i$ in terms of the initial and final partition functions $Z_i$ and $Z_f$.}
The degree of approximation is then made precise by switching from a statement that the magnitude of the left hand side of Eq \eqref{eq: rearrangedAQC} is approximately zero, to a statement that it is less than or equal to some, hopefully small, error bound,
\begin{equation}\label{eq: Error Bounded Equality}
\begin{aligned}
&D := | Z_i \bra{\psi_i}\exp\left(- \frac{H_B}{k_B T}\right)\ket{\psi_i}  \mathcal{P}\left( \phi_f | \phi_i, \gamma_i \right)  \\ &- Z_f \bra{\phi_f}\exp\left(- \frac{H_B}{k_B T}\right)\ket{\phi_f}\mathcal{P}\left(\psi_f | \psi_f, \gamma_f \right) | \leq \epsilon .
\end{aligned}
\end{equation}
It is possible to determine analytically the terms $\bra{\psi_i}\exp\left(- \frac{H_B}{k_B T}\right)\ket{\psi_i}$ and $\bra{\psi_f}\exp\left(- \frac{H_B}{k_B T}\right)\ket{\psi_f}$ in $D$ for the coherent, squeezed and cat state equalities by arranging their approximate forms.

The total error, $\epsilon$, is the sum of the initial and final factorisation errors,
\be \label{eq: total error}
\epsilon = ||\epsilon_{SB}^i||\zo{_1} + ||\epsilon_{SB}^f||\zo{_1} \ 
\ee
\flo{where $||\epsilon||_1=\Tr\left[\sqrt{\epsilon^\dagger\epsilon}\right]$ denotes the trace norm of the operator $\epsilon$.}
These factorisation errors $\epsilon_{i,f}$ capture the extent to which the exponential of the total Hamiltonian, $H_{SB}$, factorises into the exponential of effective battery and system Hamiltonians, $H_B$ and $H_S^{i,f}$, when acting on the battery states that are measured in subregions $\Pi_B^{i,f}$
(i.e. $\ket{\psi_f}$ and $\ket{\phi_f}$ respectively).
More accurately, and as derived in Appendix \ref{Appendix: Derivation of AQC}, having defined the map
\be
J_K(\zo{\sigma})=e^{-\frac{K}{2 k_B T}} \zo{\sigma} e^{-\frac{K}{2k_B T}} \, ,
\ee
the factorisation errors are defined as 
\begin{equation}\label{eq: factorisation error}
\begin{aligned}
& \epsilon_{SB}^i = J_{H_{SB}}(\I_S \otimes\ket{\psi_i}\bra{\psi_i})-J_{K_{SB}^i}(\I_S \otimes\ket{\psi_i}\bra{\psi_i}) \\
& \epsilon_{SB}^f =  J_{H_{SB}}(\I_S \otimes\ket{\phi_f}\bra{\phi_f})-J_{K_{SB}^f}(\I_S \otimes\ket{\phi_f}\bra{\phi_f}) 
\end{aligned}
\end{equation}
where $K_{SB}^i =  H_S^i\otimes\I_B +\I_S \otimes H_B$ and $ K_{SB}^f = H_S^f \otimes\I_B +\I_S \otimes H_B$.
As calculating $\epsilon$ involves exponentiating $H_{SB}$, this generally needs to be done numerically. 

\medskip

 \zo{While Eq.~\eqref{eq: Error Bounded Equality} enables the AQC to be tested, $D$ and $\epsilon$ are not always reliable indicators of the accuracy of the AQC.} For example, $D$ decreases as the transition probabilities $\P$ decrease but this need not correspond to the AQC holding more accurately. Therefore, we also introduce $1 - R$ as an alternative means of quantifying its accuracy, with $R$ the ratio of the two sides of the AQC,
\be
R := \frac{\mathcal{P}(\phi_f|\phi_i, \gamma_i)}{\mathcal{P}(\psi_i|\psi_f, \gamma_f)} \left(\exp\left(\frac{\Delta \tilde{E}-\Delta F }{k_B T}\right) \right)^{-1}\ . 
\label{eq:Rvalue}
\ee
The quantity $1 - R$ equals 0 when the AQC holds exactly. 

\section{Proposed Physical Implementation}\label{Sec: Physical Implementation}
To make concrete the quantum and thermal physics governed by the AQC we specify a particular interaction, Eq.~\eqref{eq:interaction}, between system and battery for which we expect the equality to hold and the procedure for testing it. We start by considering an `ideal' interaction that replicates the classical Crooks equality setup most closely. We then take a broader perspective and suggest alternate potentials that could be used to verify the physics quantified by the AQC.

\subsection{Position dependent level shift}\label{sec: proposal}

\begin{figure}
  \centering
\subfloat[$\mathcal{P}\left( \alpha_f | \alpha_i \exp(-\chi) , \gamma_i \right)$]{\includegraphics[width = 3.0in]{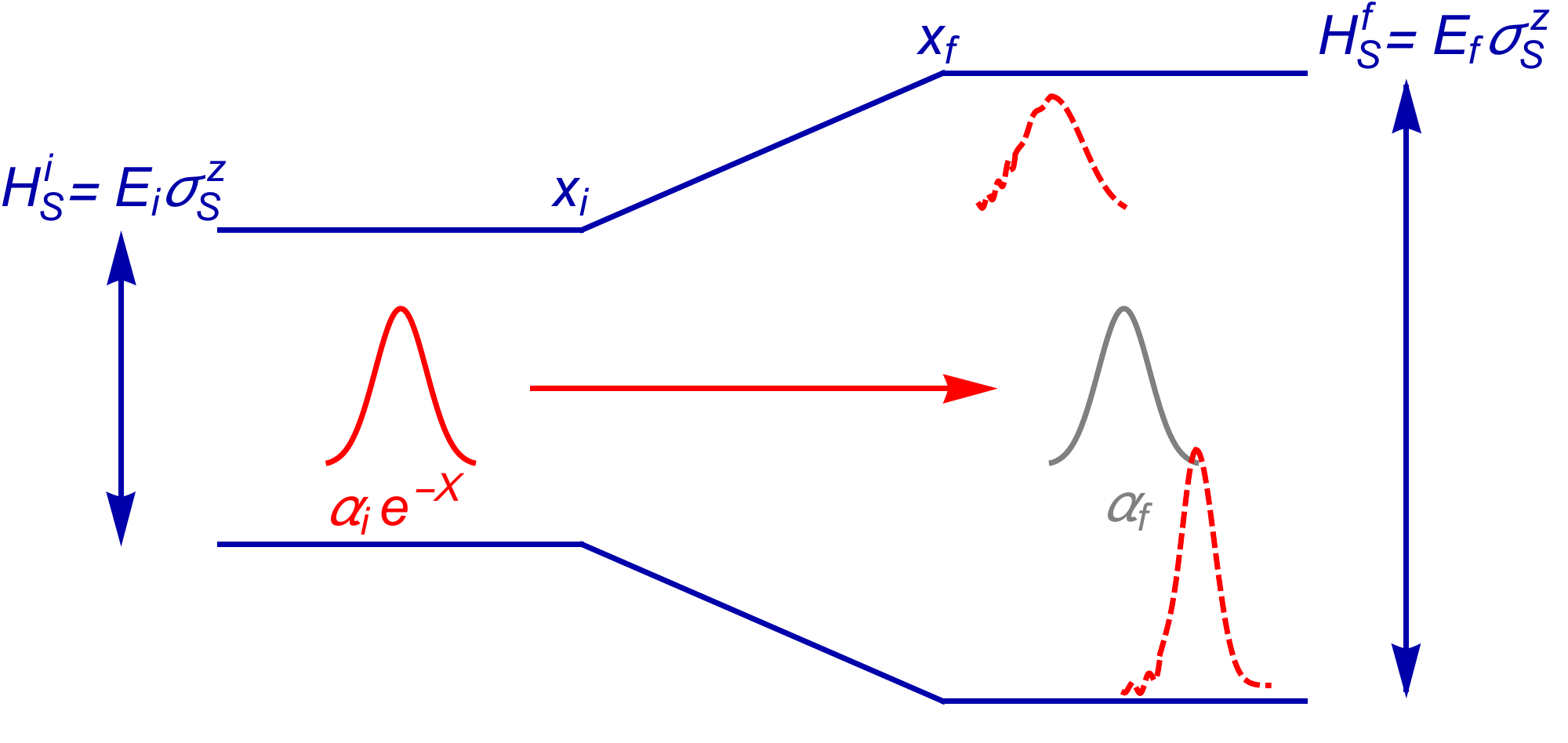}} \\
\subfloat[$\mathcal{P} (\alpha_i^* |  \alpha_f^* \exp(-\chi) , \gamma_f )$]{\includegraphics[width = 3in]{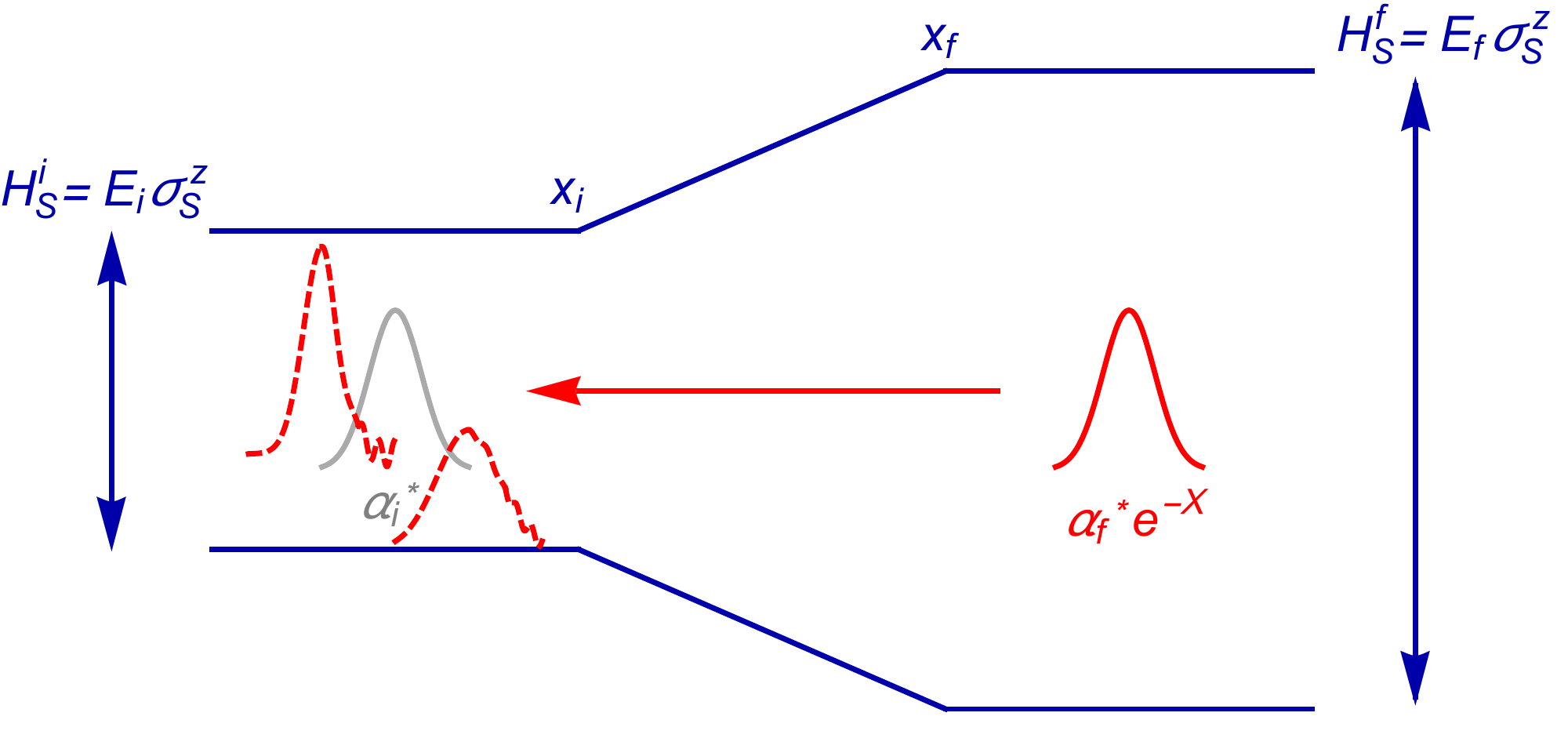}} 
\caption{Diagrammatic representation of the protocol to test the coherent state Crooks equality. The blue lines represent the ground and excited state of the system as a function of $x$. The solid Gaussians represent the coherent states that are prepared at the start of the protocols  (red) and the measurements (grey) that are performed at the end of the protocols. The dashed lines represent the evolved states. The harmonic trap that drives the evolution is centered at the midpoint of $x_i$ and $x_f$.}
\label{Fig: single protocol diagram}
\end{figure}

Following~\cite{aberg}, we consider a two level system interacting with a harmonic oscillator battery via 
\be\label{eq: level splitting interaction}
V_{SB} = \sigma^z_S \otimes E(x_B)\, ,
\ee
where $\sigma^z_S = \ket{e}\bra{e} - \ket{g}\bra{g}$ with $\ket{e}$ and $\ket{g}$ the excited and ground states of the two level system respectively. $E(x_B)$ is an energetic level-shift that depends on the position operator, $x_B$, of the oscillator. 
By choosing $E(x)$ to be constant for $x\le x_i$ and for $x\ge x_f$, two distinct effective Hamiltonians, $H_S^i$ and $H_S^f$ in Eq.~\eqref{eq:interaction}, can be realised for the two level system. The independence of the AQC on the form of $V_{SB}^\perp$ in Eq. \eqref{eq:interaction} means that the choice of the form of $E(x)$ in the region $ x_i < x < x_f$ is arbitrary and so for simplicity we assume a linear increase,
\begin{equation}\label{eq: form of level splitting}
\begin{aligned}
&E(x) =
\left\{
\begin{array}{ll}
E_i  & \ \ \ x \leq x_i \\
\frac{E_f - E_i}{x_f - x_i} (x - x_i) + E_i &  \ \ \  x_i < x < x_f \\
        E_f & \ \ \ x \geq x_f  \ \ \ \ .
\end{array} 
\right.
\end{aligned}
\end{equation}
This energy level splitting profile is shown in Fig. \ref{Fig: single protocol diagram}. Without loss of generality, we chose to center the interaction region around $x= 0$ such that $|x_i| = |x_f|$. 

\medskip

\medskip

In order to measure the transition probability, $\mathcal{P}(\phi_f|\phi_i, \gamma_i)$, for the forwards process one has to be able to implement the following three steps:
\begin{enumerate}
\item The battery is prepared in a state $\ket{\phi_i}$ that is localised such that $\bra{\phi_i} x_B \ket{\phi_i} < x_i$
and the system is prepared in a thermal state with respect to its effective Hamiltonian $H_S^i = E_i \, \sigma_S^z$.

\item The system and battery are allowed to evolve for some time $\tau$. The time $\tau$ is chosen such that the final battery wavepacket is localised in the region beyond $x_f$ in order to ensure that the effective system Hamiltonian has changed to $H_S^f = E_f \sigma_S^z$.

\item The battery is measured to determine whether its evolved state is $\ket{\phi_f}$.
\end{enumerate}
The forwards transition probability, $\mathcal{P}(\phi_f|\phi_i, \gamma_i)$, is the relative frequency with which the battery is found in the state $\ket{\phi_f}$ when the steps 1-3 are repeated a large number of times. \change{We discuss how the preparation and measurement steps might be practically implemented in a trapped ion system in Section IV B.} 

The procedure to obtain the reverse process transition probability, $\mathcal{P}(\psi_i|\psi_f, \gamma_f)$, is entirely analogous. The battery is prepared in the state $\ket{\psi_f}$ in the region beyond $x_f$ and the system is prepared in a thermal state with respect to $H_S^f= E_f \sigma_S^z$. A final measurement, after evolving for time $\tau$, determines whether the battery is in state $\ket{\psi_i}$.

A sketch of the test for the coherent state Crooks equality is shown in Fig. \ref{Fig: single protocol diagram}. The procedure for the coherent, squeezed, and cat state equalities differ only in the choice of preparation and measurement states, as specified by the table in \ref{Table}.  

\medskip

\subsection{Numerical calculations of the error bounds}\label{Sec: Results}

\zo{Here we numerically determine the error in the approximate equalities. While coherent, squeezed and cat states have support outside of the regions $x < x_i$ and $x > x_f$,
and so the AQC is not obeyed perfectly, we find that there are regimes where the errors fall below the expected experimental error margins. In such regimes, which we characterise below, the AQC is exact for all practical purposes.}

\begin{figure}
\centering
\subfloat[Coherent State and Squeezed State Equalities]{\includegraphics[width = 3.5in]{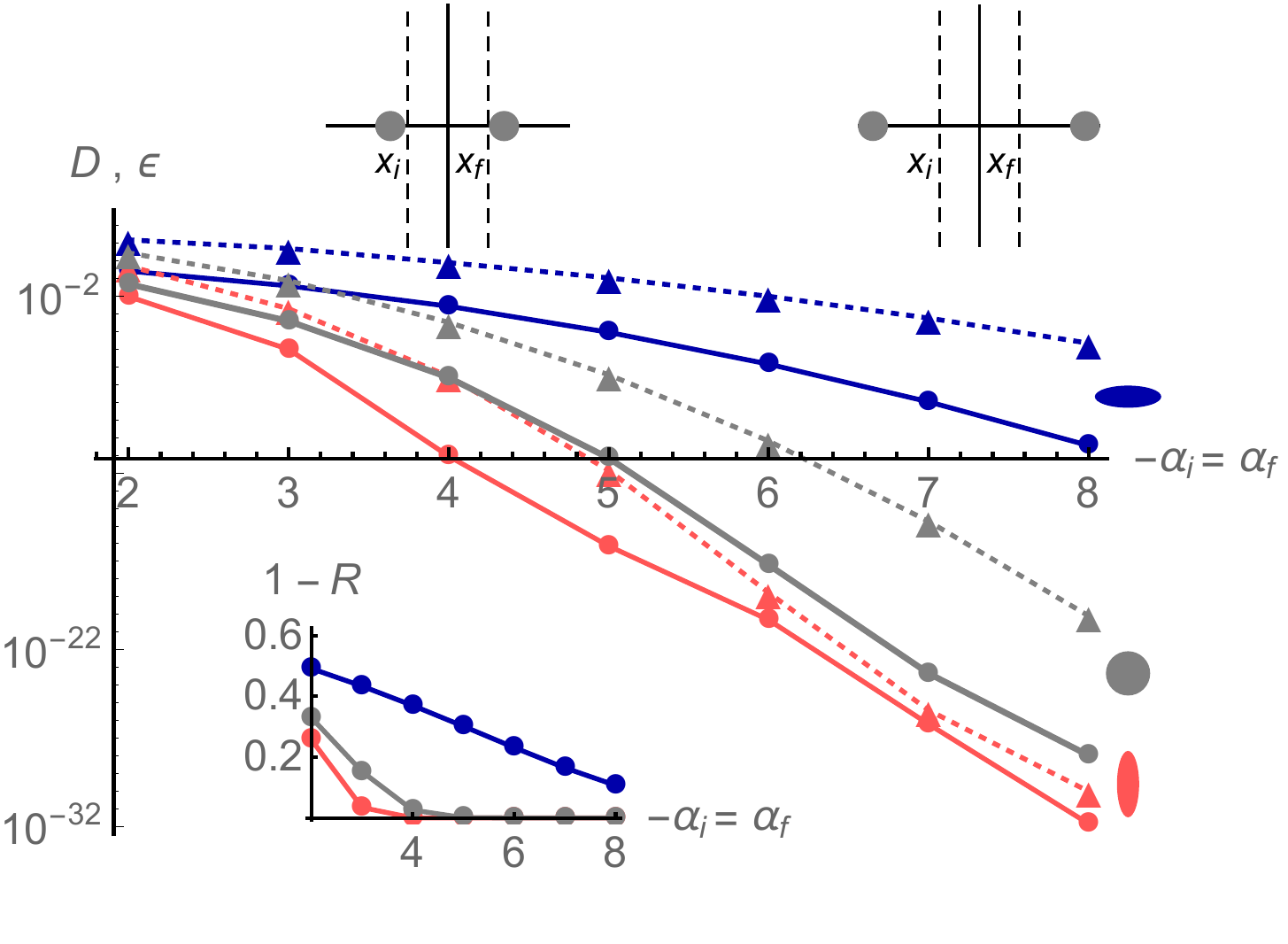}} \\
\subfloat[Cat State Equality]{\includegraphics[width = 3.5in]{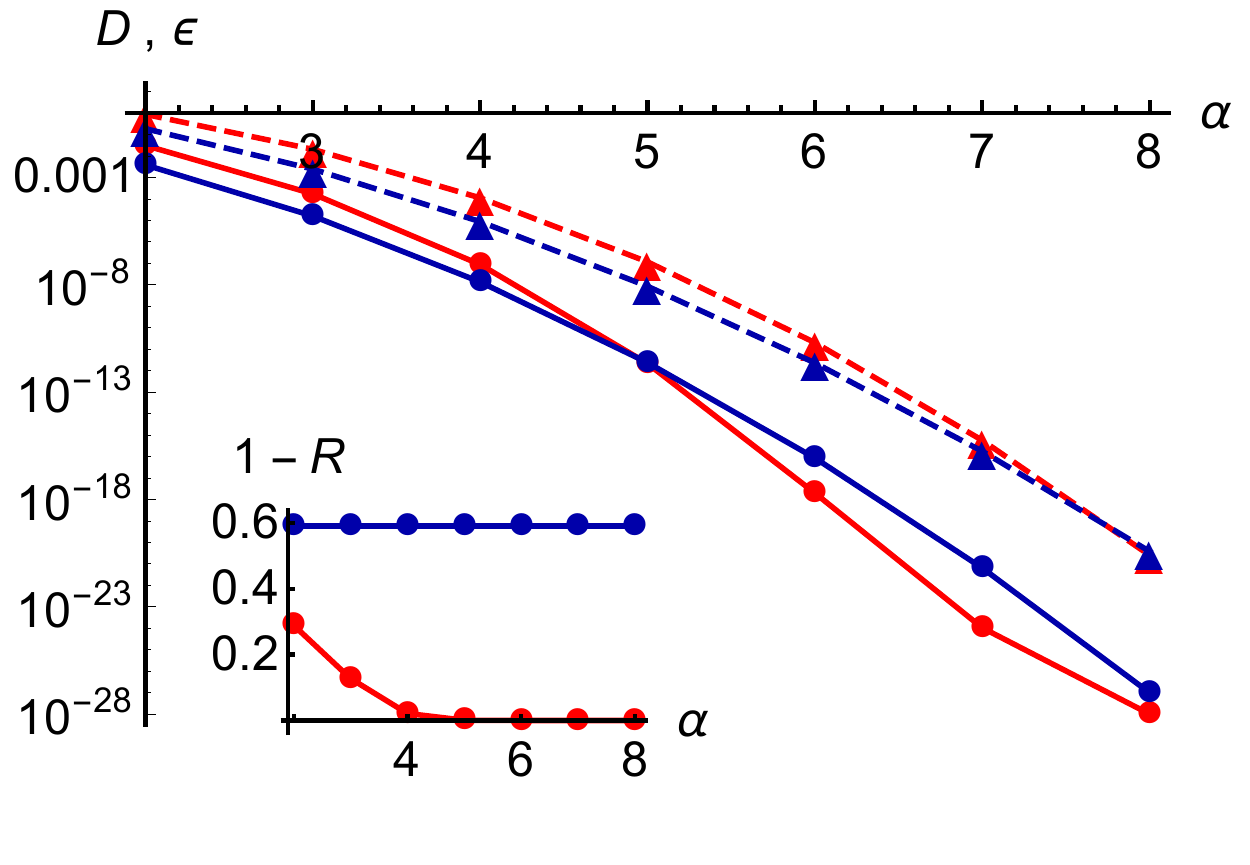}} 
\caption{The solid and dashed lines of the main plots show the error measures $D$ and $\epsilon$ respectively, and the insets plot the error measure $1-R$, for evolution under the total Hamiltonian with the position dependent level splitting interaction and a harmonic oscillator battery. In (a) these are plotted as a function of $-\alpha_i = \alpha_f$ with the coherent state Crooks equality data shown in grey and squeezed state Crooks equality data shown in blue  ($r = -1$) and red ($r=1$) (as defined in the table in \ref{Table}). In (b) the data for cat states prepared and measured straddling the interaction, $\ket{\psi_i} = \ket{\phi_f} \propto \ket{\alpha} + \ket{-\alpha}$, is shown in blue and for cat states prepared and measured on a single side of the interaction region, $\ket{\psi_i} \propto \ket{-\alpha} + \ket{-(\alpha+1)}$ and $\ket{\phi_f} \propto \ket{\alpha} + \ket{\alpha+1}$, is shown in red. In this simulation we have used the following parameters: $x_i = -4$, $x_f=4$, $\hbar \omega = 1$, $E_i= 1$, $E_f = 2$ and $k_B T = 1$. The displacement parameters, $\alpha_{i,f}$, and the positions $x_{i}$ and $x_f$, are given in units of dimensionless position $X = \sqrt{\frac{m \omega}{2 \hbar }} x$.}
\label{Fig: Equalities Hold}
\end{figure}

We simulated the full quantum state evolution of the system and harmonic oscillator battery under the total Hamiltonian with the position dependent level splitting, $E(x)$, and used this to find the error measures $D$ and $R$, as defined in Eq.~\eqref{eq: Error Bounded Equality} and Eq.~\eqref{eq:Rvalue}. We further numerically calculated the error bound $\epsilon$, defined in Eq.~\eqref{eq: total error}. As expected we find that $D$ for the coherent, squeezed and cat states is less than $\epsilon$, for a wide range of simulated parameters ($k_B T$, $\omega$, $\alpha_i$, $\alpha_f$, $r_i$, $r_f$, $\beta_i$, $\beta_f$, $E_{i}$, $E_f$, $x_i$, $x_f$). Fig. \ref{Fig: Equalities Hold} presents some of these results.

For the coherent states and squeezed states, with \flo{real $\alpha_i$ and real $\alpha_f=- \alpha_i$}, the inset of Fig.~\ref{Fig: Equalities Hold}a shows that $1-R$ tends to 0 as $\alpha_i$ is increased. \zo{As is shown in red and blue respectively, for a given $\alpha_i$ squeezing the position variance of the prepared and measured states decreases $1- R$; while, squeezing the momentum variance increases $1- R$.}
\flo{Coherent states with also a complex displacement $\alpha$ corresponding to finite momentum result in similar behaviour, as we have explicitly verified.}
For the cat state equality, we firstly consider final states of the form $\ket{\psi_i} \propto \ket{-\alpha} + \ket{-(\alpha +1)}$ and $\ket{\phi_f} \propto \ket{\alpha} + \ket{\alpha+1}$. On doing so, as is shown by the red line in the inset of Fig. \ref{Fig: Equalities Hold}b, we find that, as with the coherent state case, $1-R$ tends to 0 as $\alpha$ is increased. However, for the cat states $\propto\ket{\alpha} + \ket{-\alpha}$, we find that $1-R = 0.59$ for all $\alpha$, as shown by blue line. In contrast to the other states we simulated that are initially displaced from the origin, these cat states are symmetrically delocalised such that the average battery position is 0. In particular, there may be a significant probability to find the battery in both regions, $x < x_i$ and $x > x_f$. 


We conclude that decreasing the overlap of the prepared and measured battery states from the interaction region increases the accuracy of the AQC. Moreover, the convergence to the exact limit is reached quickly. \zo{For example, we see in Fig.~\ref{Fig: Equalities Hold}a that for coherent states with $|\alpha_i|=|\alpha_f| \geq 6$ the error measures are very small, i.e. $D,\,  \epsilon,\,(1-R) < 10^{-6}$.} Such discrepancies are lower than experimental error margins and consequently the approximate nature of these equalities can effectively be disregarded when the preparation and measurement states are chosen appropriately.

The behaviour of $(1-R)$ in Fig.~\ref{Fig: Equalities Hold} makes physical sense. Fundamentally, the approximate nature of the AQC arises because, by considering the autonomous evolution under a time-independent interacting Hamiltonian, the system never strictly has a well-defined local Hamiltonian. The crux of the issue is the battery Hamiltonian induces evolution between subregions with different effective Hamiltonians, $H_S^i$ and $H_S^f$. As a result, the thermal state of the system at the start of the forward and reverse protocols is not well defined. However, when the battery is initially prepared with support in a single region far from the interaction region the system can nonetheless properly thermalise with respect to its initial Hamiltonian. In this limit the AQC is exact.

\subsection{Alternative position dependent level splittings}\label{sec: alternative potentials}

Thus far we have specified one particular choice in $E(x)$, Eq.~\eqref{eq: form of level splitting}, to induce an effective change in system Hamiltonian. Here we outline how alternative choices in $E(x)$ can also be used to probe the physics quantified by the AQC.

In Eq.~\eqref{eq: form of level splitting} we took $E(x)$ to be constant in the regions $x \leq x_i$ and $x \geq x_f$ to ensure that the initial and final Hamiltonians are well-defined and so the AQC holds to a high degree of approximation. However, there is more flexibility as to the choice of $E(x)$ if, rather than aiming to directly verify the AQC, we instead focus simply on detecting the quantum deviation from the classical Crooks equality that the AQC predicts.

The deviation of the AQC from the classical Crooks equality is encapsulated by the prefactor \zo{$q := \Delta \tilde{E}/W_q$, as per Eq.~\eqref{eq: q factor}}, that appears in the term $\exp\left(q \frac{ W_q}{k_B T}\right)$ of the predicted ratio of transition probabilities.
\zo{The prefactor $q$ can always be inferred from the transition probabilities using
\begin{equation}\label{eq:qdef}
\begin{aligned}
q = \frac{k_B T}{W_q} \zo{\ln}\left(
\frac{\mathcal{P}(\phi_f|\phi_i, \gamma_i)}{\mathcal{P}(\psi_i|\psi_f, \gamma_f)} \exp\left(\frac{\Delta F}{k_B T}\right) \right) \ ;
\end{aligned}
\end{equation} 
however, when the AQC does not hold exactly, it will deviate from $\Delta \tilde{E}/W_q$.} 

\begin{figure}
\centering
{\includegraphics[width = 3.4in]{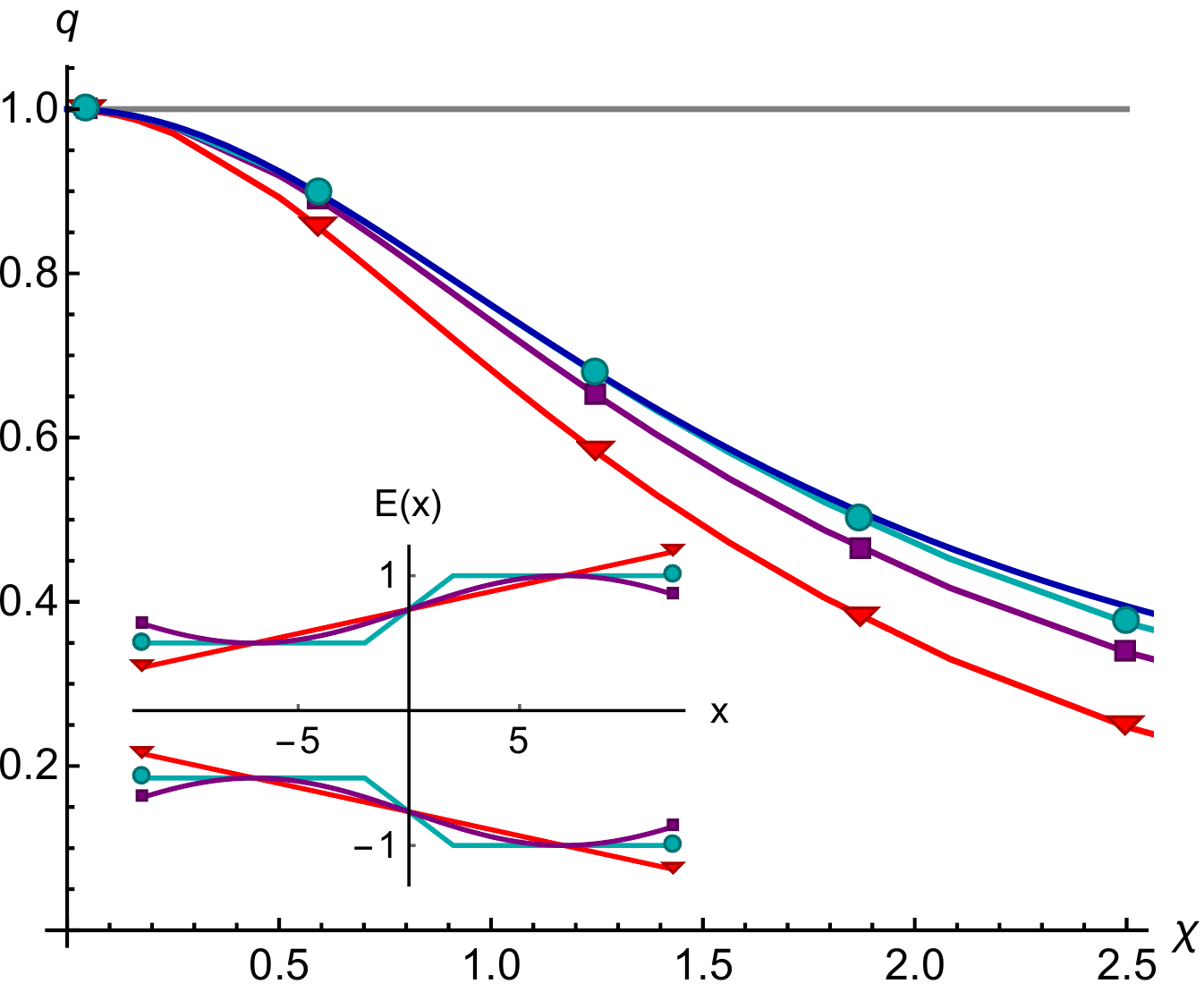}}
\caption{The predicted quantum deviation $q$ deduced from the transition probabilities between coherent states, using Eq.\eqref{eq:qdef}, for different position dependent level splittings, $E(x)$. The turquoise circles show the inferred $q$ for the usual `flat ends' potential, the purple squares show $q$ for a $\sin(x)^2$ potential, and the red triangles show $q$ for a purely linear potential. The blue line is the predicted analytical form of $q = \frac{\tanh(\chi)}{\chi}$ (also sketched in blue in Fig.~\ref{Fig: squeeze q factor}). The grey line indicates the classical limit in which $q = 1$. The inset plots the level splittings simulated. In this simulation we have chosen the following physically plausible parameters, $x_i = -2$, $x_f=2$, $E_i= 1$MHz, $E_f = 2$MHz, $\alpha_{i} = -5$ and $\alpha_{f} =4$. The displacement parameters $\alpha_{i,f}$ and the positions, $x_{i}$ and $x_f$, are given in units of dimensionless position.}
\label{Fig:Varying Potentials}
\end{figure} 

We numerically simulated the evolution of a coherent state for different choices of $E(x)$ and used the obtained transition probabilities to infer $q$, which we plot in Fig.~\ref{Fig:Varying Potentials}. The analytic form of $q$ for coherent states, Eq.~\eqref{eq: q factor coherent}, is indicated by the dark blue line. As shown by the turquoise circles, we find for $E(x)$ with the `flat ends', Eq.~\eqref{eq: form of level splitting}, that $q$ closely replicates the predicted analytic form. The change in effective Hamiltonian from $E_i \sigma_S^z$ to  $E_f \sigma_S^z$ can be approximately replicated by using a $\sin(x)^2$ function and choosing $\alpha_i$ and $\alpha_f$ to sit in the troughs and peaks of the potential where the effective Hamiltonian is approximately constant. \zo{As shown in purple, for this choice in $E(x)$ the behaviour of $q$ reasonably closely replicates the analytic expression.} If we abolish the flat ends entirely and instead enact a change in effective Hamiltonian using a solely linear potential then, as shown by the red triangles, we find that $q$ deviates further from the predicted analytic form but is still of a similar functional form.

As such, we conclude that it should be possible to detect quantum deviations using sinusoidal and linear potentials as well as plausibly other potentials that have yet to be explored. This flexibility as to the choice of $E(x)$ opens up many possible realisations to verify the physics quantified by the AQC.
\medskip

\section{Experimental Outline: Trapped Ion and AC Stark Shift}

To make the proposal more concrete we will now outline a potential experimental implementation utilising an ion confined to a linear Paul trap. A pair of internal electronic levels and the elongated phonon mode represent the system and battery respectively.

For an experimental verification of the AQC to be convincing and practical we have the following six conditions.

\begin{enumerate}
\item Practically, we require forwards and reverse transition probabilities of a measurable order of magnitude, \flo{{\it i.e.}}
\be\begin{array}{rcccll}
10^{-2} & \lesssim & \mathcal{P}(\phi_f|\phi_i, \gamma_i) & \lesssim & 1 - 10^{-2} \ & \mbox{and}\\
10^{-2} & \lesssim & \mathcal{P}(\psi_i|\psi_f, \gamma_f) & \lesssim & 1 - 10^{-2}\ .
\end{array}
\nonumber\ee
\item  To practically simulate a non-trivial thermal system state we require non-negligible populations of both the ground and excited states. This requires the initial and final level splittings, \zo{$2 E_i$ and $2 E_f$} respectively, to be a similar order of magnitude to the temperature, {\it i.e.}
\be
\zo{2 E_i} \approx k_B T   \approx \zo{2 E_f}\ .\nonumber
\ee
\item To non-trivially test the AQC we require a moderately strong system-battery interaction. This is ensured as long as the change in the splitting of the system energy levels is of a similar order of magnitude as the trap frequency, \flo{\it i.e.}
\be
| \zo{2 E_f - 2 E_i}| \approx \hbar \omega\ .\nonumber
\ee
\item To cleanly test the AQC, the decoherence and heating rates of the battery need to be considerably less than the trap frequency. If long-lived internal levels are used to represent the two-level system then the primary constraint is the heating rate of the phonon mode battery, $\nu_{\mbox{\tiny therm}}$, and thus we need,
\be
\omega \gg \nu_{\mbox{\tiny therm}}\ .\nonumber
\ee
\item The interaction needs to be designed in such a way that the ion can be prepared in a state comfortably in the regions $x<x_i$ and $x > x_f$ with minimal overlap with the interaction region, $x_i < x < x_f$.



\end{enumerate} 

The simplest means of inducing our proposed Hamiltonian with the level splitting specified by Eq.~\eqref{eq: form of level splitting} would be to use a pair of Zeeman split internal electronic levels and a position dependent magnetic field, $B(x) \propto E(x)$. However, it is challenging to obtain the required magnetic field environment for currently obtainable ion heights and corresponding motional heating rates.

    

\begin{figure}
  \centering
{\includegraphics[width = 3.4in]{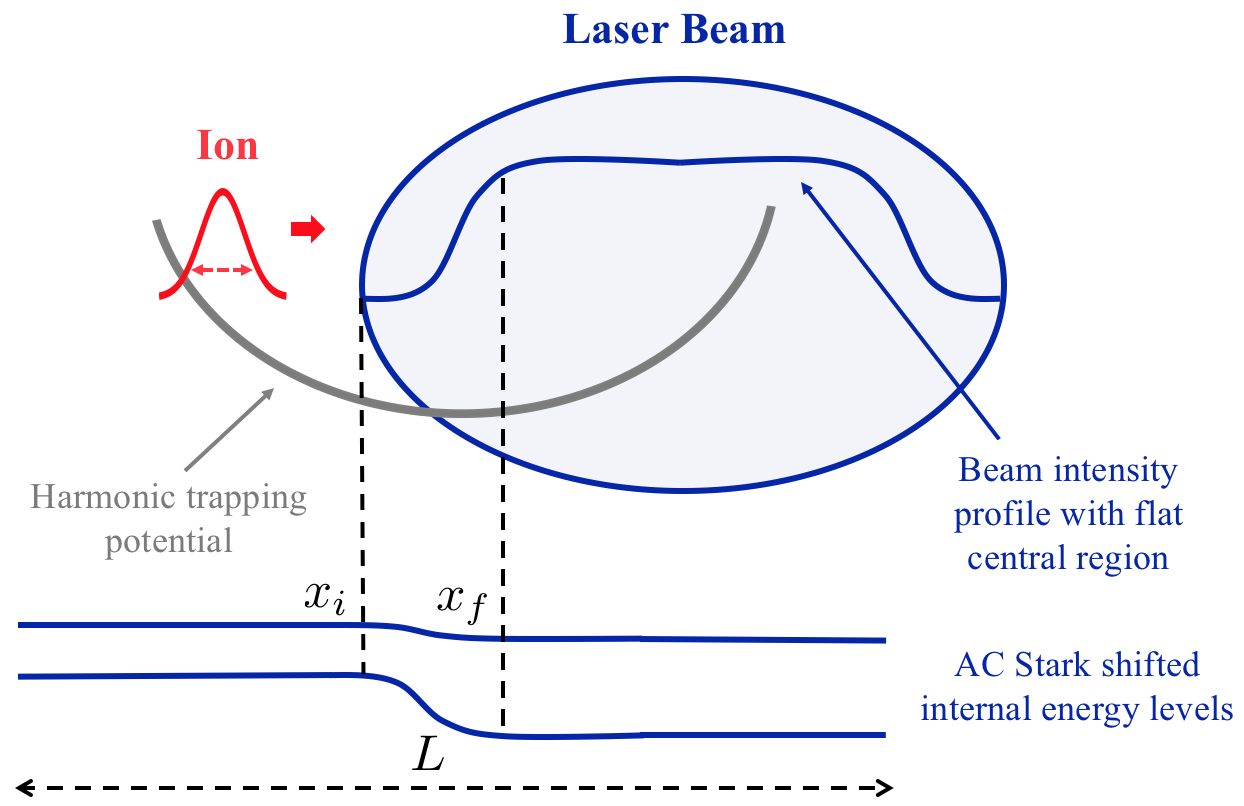}}
\caption{Sketch of trapped ion proposed implementation. The ion (red wavepacket) is at the start of the evolution stage of the forwards protocol. It is displaced in an elongated trap (grey line) that drives its evolution.  An off-resonance laser beam propagates perpendicularly across one side of the trap (blue oval). The laser has an intensity profile that is sloped on the edges and flat through the center (blue line in oval). The trapped ion experiences a position dependent AC Stark shift (pair of blue lines) as it travels autonomously through the laser beam. The red dashed line indicates the spread of the ion and the black dashed line the length of the trap. The preparation and measurement stages are sketched in Fig.~\ref{Fig: prepandmeasure}.}
\label{Experimental Apparatus}
\end{figure} 

We instead advocate using an AC Stark shift, generated by an off resonant laser field, to induce a position dependent level splitting. \zo{While the magnetic field realisation is perfectly autonomous, strictly speaking this implementation allows the exchange of photons (and thus energy) between the ion and the light field, and therefore is not perfectly autonomous. However, the energy exchange quickly decays for increasing detuning and for experimentally achievable Rabi-frequencies and detunings the system can nonetheless be considered autonomous for all practical purposes.}

To replicate Eq.~\eqref{eq: form of level splitting} as closely as possible we require the laser beam to have an intensity profile with a flat region corresponding to $x > x_f$. In the region $x < x_i$ there is no laser beam and the Hamiltonian is $H_S^i$, while in the region $x>x_f$ a flat laser beam profile induces a constant AC Stark shift realising $H_S^f$. In the region  $x_i < x < x_f$ the tail of the laser beam changes the Hamiltonian from $H_S^i$ to $H_S^f$. In Fig.~\ref{Experimental Apparatus} we sketch this for a smoothed top hat potential. However, in general the tails of the intensity profile need not be smoothed as there are no constraints on the interaction in the region $x_i$ and $x_f$. 

The AC Stark shift typically changes the average energy of the system as well as the splitting between the energy levels. This change in average energy and the smoothed change in gradient at $x_i$ and $x_f$ (as shown in Fig. \ref{Experimental Apparatus}) requires a minor change to $V_{SB}$ as specified in Eq.~\eqref{eq: level splitting interaction} and \eqref{eq: form of level splitting} but makes no pertinent difference to the underlying thermal physics being investigated.

\medskip

While we focus on the tapered beam implementation to maintain a closer resemblance to the classical Crooks equality setup, it is also possible, as discussed in Section~\ref{sec: alternative potentials}, to detect the quantum deviation predicted by the AQC using a level splitting that varies sinusoidally or linearly with position. An AC Stark shift that varies sinusoidally with position could be realised using a standing wave generated by two counter propagating lasers~\cite{schmidtkaler,spinheatengine}. A magnetic field that increases strength linearly with position $B(x) \propto x$ could be used to realise a Zeeman shift that increases linearly with position~\cite{magneticfieldgradient1,magneticfieldgradient2,magneticfieldgradient3}.

\subsection{\zo{Experimental Parameters}}

To support the plausibility of our proposal we present a set of potential parameters that satisfy the five requirements listed above and could as such enable the AQC to be verified. 

We propose using a single $^{171}$Yb$^{+}$ ion trapped in a linear Paul trap\footnote{The radial trapping strength must be large enough to freeze out the radial force on the ion that is induced by the magnetic component of the laser~\cite{barnett}.} with an axial secular frequency of 0.3 MHz~\cite{hensingerions}. Such secular frequencies typically experience heating rates of the order of 40 phonons per second~\cite{heatingrate}, thereby satisfying requirement 4.
Requirements 2 and 3 limit us to a pair of internal levels with a separation of the order of MHz and thus we are constrained to using hyperfine levels. 
We propose using the F = 1, M$_{\mbox{\tiny F}} = 1$ and M$_{\mbox{\tiny F}} = -1$ levels of the $^{2}$S$_{1/2}$ ground state. 
A magnetic field can be applied to lift the degeneracy of the F = 1 manifold~\cite{hensingerions} and set \zo{$E_{i}$} with typical splittings ranging between approximately 0.5 MHz and 10 MHz. 

We suggest using a laser field with an intensity of  $4.5$ Wmm$^{-2}$ that is red detuned from the 369nm $^{2}$S$_{1/2}$ - $^{2}$P$_{1/2}$ transition by $0.1 \times 10^{14}$Hz. Using the methods and data from~\cite{starkshift1,starkshift2}, this is expected to induce an AC Stark shift on the M$_{\mbox{\tiny F}} = 1$ and M$_{\mbox{\tiny F}} = -1$ states of -0.2 MHz and -1.2 MHz respectively.

The position variance of $^{171}$Yb$^{+}$ for an axial secular frequency of 0.3 MHz is $0.01 \ \mu m$~\cite{trappedionsreviewLeibfried}. Therefore, to satisfy requirement 5, a top hat potential with the size of the `flat' region being greater than $0.05 \ \mu$m is required. There is some flexibility in how accurately the beam profile needs to be shaped as our simulations indicate that $1-R$ scales linearly\footnote{The exact constant of proportionality depends on the temperature, the magnitude of the Stark shift and the battery wavepacket parameters; however, it is generally of the order of $1 - R \approx \frac{\delta I}{I}$ where $\delta I$ is the change in laser intensity over a distance of $0.1 \, \mu$m and $I$ is the maximum laser intensity in that region. The sensitivity to an intensity gradient is lower for higher temperatures, smaller Stark shifts and battery states that are prepared comfortably outside the interaction region.} with a gradient across the ideally `flat' region of the beam.

Requirement 1 gives rise to a trade off between the spatial scale on which the beam can be engineered and the temperature regime that can be probed. As the distance $x_f - x_i$ is the minimum the ion must travel to realise the change in Hamiltonian $H_S^i$ to $H_S^f$, this sets the minimum possible displacement of the initial and final states. When the ion oscillates over large distances, its position variance is relatively small. In this limit, it is more challenging to choose the initial and final ion states such that there is a significant overlap between the measured states and the evolved states in both the forwards and reverse processes. It is easier to ensure there is a significant overlap in the high temperature regime because the pair of states prepared in the reverse process are approximately the time reverse of the pair of states in the forwards process. Our simulations indicate that the $k_B T \approx 25 \, \hbar \omega$ regime can be probed with $x_f - x_i = 1 \mu$m. The experiment would be feasible with a larger $x_f - x_i$; however, the shorter the spatial scale on which the ion oscillates, the lower the temperature regime it would be possible to probe and the greater the $\chi$ induced quantum deviations can be detected\footnote{It is possible to realise standing waves and linear magnetic field gradients on the scale of $10^{-2} \mu m$~\cite{magneticfieldgradient3,schmidtkaler}. Thus the small spatial scales of these alternative approaches could enable larger values of $\chi$ to be more easily probed.}. 

\subsection{Experimental Techniques}

The procedure outlined in \ref{sec: proposal} is realisable for trapped ions using the following currently available experimental techniques.  

\paragraph*{Preparation of two-level system.} 
A thermal state of the internal electronic energy levels can be modeled using `pre'-processing~\cite{superconductingqubitdeamon}. Specifically, a thermal `ensemble' of $N = N_g + N_e$ ions can be modeled by running the protocol on a single ion $N_g$ and $N_e$ times (where $N_e/N_g = \exp(-\zo{2 E_i}/ k_B T)$) with the ion initialised in the ground state and excited states respectively. The ion can be prepared in the ground state or excited state by using an appropriate sequence of laser and microwave pulses. Simulating the thermal state in this way makes it straightforward to ensure that we are in the low temperature limit with respect to the trapping frequency.

\begin{figure}[H]
\centering
\subfloat[Preparation]{\includegraphics[width = 3.4in]{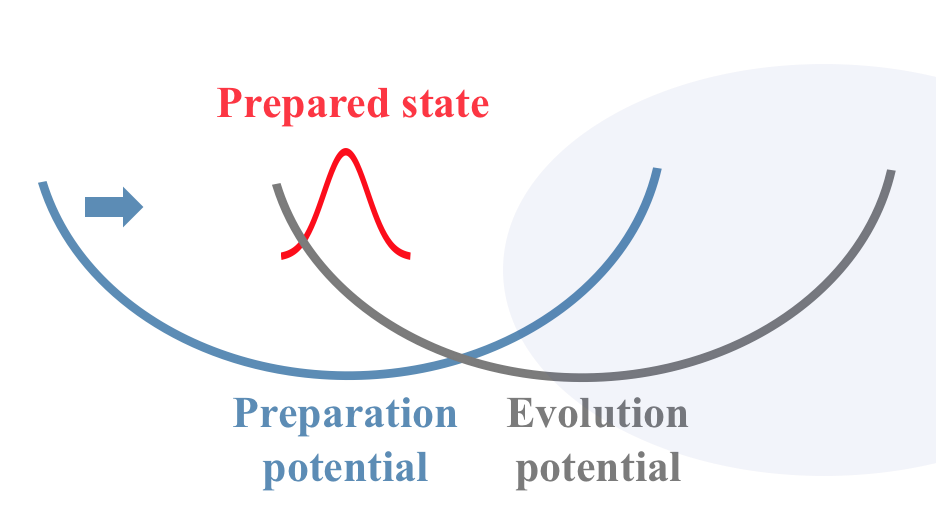}} \\
\subfloat[Measurement]{\includegraphics[width = 3.4in]{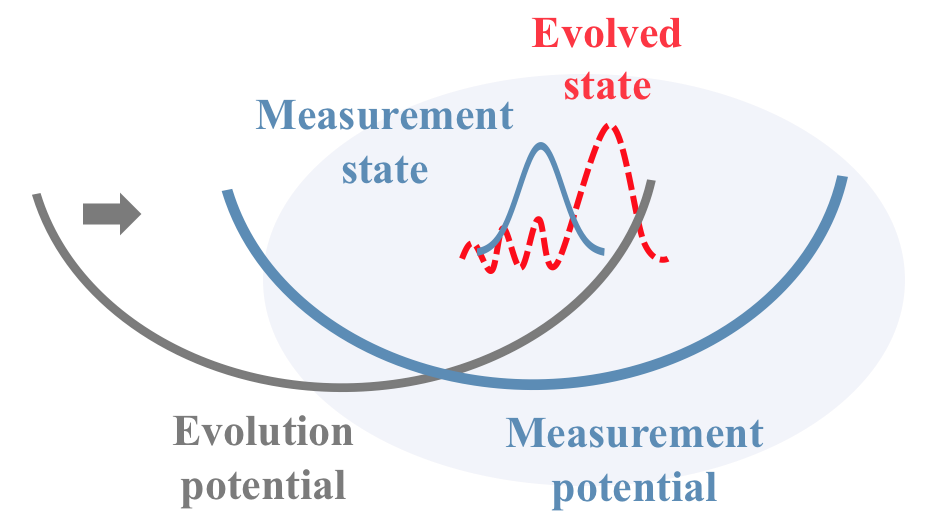}} 
\caption{Sketch of preparation and measurement stages of the forwards protocol of the coherent state trapped ion implementation. (a) To prepare a coherent state the ion is first prepared in the motional ground state of a harmonic trapping potential, shown here in blue. The potential center is then shifted non-adiabatically resulting in the effective displacement of the ion with respect to the new trapping potential, shown here in dark grey. (b) To find the overlap between the evolved state of the ion (the dashed red wavepacket) and some test coherent state (the blue Gaussian), the trap potential is first shifted such that were the evolved state precisely in the test state, then the evolved state would be in the ground state with respect to the new trapping potential. This shifted trapping potential is shown in blue. A projective measurement is then performed to find the overlap between the evolved state and the ground state of the new trapping potential. In both figures the blue oval in the background is the laser beam that induces the AC Stark shift.}
\label{Fig: prepandmeasure}
\end{figure}

\paragraph*{Preparation of oscillator states.} The first step to generate coherent, squeezed and cat states is to prepare the ion in the motional ground state using sideband cooling. A coherent state can then be generated by shifting the trap center non-adiabatically resulting in the effective displacement of the ion with respect to the new trapping potential~\cite{trappedionsreviewLeibfried} (this is sketched in Fig.~\ref{Fig: prepandmeasure}). A squeezed displaced state can similarly be generated from the motional ground state by first generating a squeezed vacuum state using a non-adiabatic drop in the trap frequency and then displacing it using the non-adiabatic shift of the trap center~\cite{trappedionsreviewLeibfried}. 

Cat states can be generated using laser pulses that entangle the internal electronic states and motional states of the ion~\cite{trappedionsreviewLeibfried}. Recently a cat state separated by 259nm has been achieved~\cite{catstates} and with this apparatus it would be possible to generate states separated by $\mu m$ as required here.

\paragraph*{Measurement of oscillator states.}
A measurement to determine the overlap of the final state of the phonon modes, $\ket{\psi_{\mbox{\tiny final}}}$, with some test state, $\ket{\psi_{\mbox{\tiny test}}}$, can be achieved by performing the inverse of the relevant preparation process, \zo{$U_{B}^{\mbox{\tiny test}\dagger}$}, and then measuring to determine whether the battery is in the motional ground state $\ket{0}$, i.e. the overlap $\braket{\psi_{\mbox{\tiny test}}|\psi_{\mbox{\tiny final}}} = \bra{0}  \zo{U_{B}^{\mbox{\tiny test}\dagger}} \ket{\psi_{\mbox{\tiny final}}} $.
For example, to determine the overlap of some final state with the coherent state $\ket{\alpha_{\mbox{\tiny test}}}$ we would have $  \zo{U_{B}^{\mbox{\tiny test}\dagger}} = D(\alpha_{\mbox{\tiny test}})^\dagger = D(-\alpha_{\mbox{\tiny test}})$ (this is sketched in Fig.~\ref{Fig: prepandmeasure}).\footnote{There is some flexibility on how fast the measurement needs to be performed. The inverse process $U_{\mbox{\tiny test}}^\dagger$ needs to be performed at time $\tau$ or when the ion returns to the same place after another complete rotation, i.e. $(2n +1) \tau \ \ \forall n \, \in \mathbb{Z}^+$. The measurement to determine whether the ion is in the motional ground state need only be performed before there is a substantial probability for the state to be disturbed by decoherence or heating.}

A filtering scheme can be used to perform a projective measurement on a single ion that answers the binary question `is the phonon in the state $n = m_{\mbox{\tiny test}}$?'~\cite{huberjar,huberthesis}. Choosing $ m_{\mbox{\tiny test}} = 0 $, this method can be used to determine whether the ion is in the motional ground state. The filtering scheme uses the dependence of the Rabi frequency of the red and blue sidebands on phonon number.

\section{Conclusions and Outlook}\label{protocol diagram}

The AQC is derived from a simple set of physical principles and in virtue of this the coherent, squeezed and cat state Crooks equalities are both natural and general. Specifically, the derivation of the AQC only assumes that the system and battery evolve under a microscopically energy conserving and time reversal invariant unitary and that the system and battery obey a `factorisability' condition. (This condition, defined in Appendix \ref{Appendix: Derivation of AQC}, characterises the extent to which the system and battery are independent subsystems at the start and end of the protocol.) Given that these are a natural set of assumptions, shared with other derivations of fluctuation theorems, it is perhaps intriguing that for coherent states, which are often viewed as the most classical of the motional quantum states, we find the AQC can be written in the satisfyingly compact form of Eq.~\eqref{eq: coherentstaterewrite}. \coh{The dependence of the coherent state Crooks equality on the thermal de Broglie wavelength is a non-classical feature, arising from coherent superpositions between energy eigenstates.}

The coherent, squeezed and cat state Crooks equalities can be viewed as setting out quantum corrections to the classical Crooks equality. In particular, the coherent state Crooks equality is its lowest order extension to quantum states. The coherent state equality effectively
reduces to the classical Crooks equality in the classical limit but by increasing the ratio of quantum fluctuations to thermal fluctuations it is possible to smoothly interpolate from the classical to the quantum regime. In the general quantum case the probability for the reverse process is greater than expected classically and correspondingly irreversibility is apparently softened. The squeezed state and cat state Crooks equalities are higher order extensions to the classical Crooks equality. The equalities are correspondingly more complex and there are regimes in which irreversibility is apparently strengthened. It would be interesting to investigate whether the squeezed and cat state equalities can be written in a more compact form by relating them to appropriate thermal wavelengths. 

Our proposal \zo{can be seen to provide a means to experimentally test} resource theories for quantum thermodynamics.
As in the protocol the system starts in a thermal state and energy is globally conserved, the operation on the battery is a thermal operation for a system with a changing Hamiltonian~\cite{alvaro}. Consequently, the proposed experiment allows us to probe coherent features of the resource framework, which is noteworthy as these frameworks~\cite{resourcetheory1,resourcetheory2, completestateinterconversion} have, with the exception of a few recent developments~\cite{Cooperpair, algorithmiccooling}, remained abstract. 


More broadly, we have taken a highly mathematical result from within the quantum information theoretic approach to thermodynamics and both explored its physical content and an experimental implementation. We hope our research encourages more such attempts to physically ground recent quantum thermodynamics theory results. 

\bigskip

\acknowledgements
The authors thank Johan \AA berg for comments on a draft, Tom Hebdige, Erick Hinds Mingo and Chris Ho for numerous helpful discussions and Stephen Barnett for highlighting the radial force induced on the ion by an off resonant laser. This research was supported in part by the COST network MP1209 “Thermodynamics in the quantum regime.” ZH is supported by  Engineering and Physical Sciences Research Council Centre for Doctoral Training in Controlled Quantum Dynamics. SW is supported by the U.K. Quantum Technology hub for Networked Quantum Information Technologies, EP/M013243/1. DJ is supported by the Royal Society. JA acknowledges support from Engineering and Physical Sciences Research Council, Grant EP/M009165/1, and the Royal Society. FM acknowledges support from Engineering and Physical Sciences Research Council, Grant EP/P024890/1.

\bibliographystyle{unsrtnat} 
\bibliography{refs}

\appendix

\onecolumngrid

\section{Derivation of the AQC}\label{Appendix: Derivation of AQC}

The AQC is derived in~\cite{aberg} by combining two properties that we will call `global invariance' and `factorisability'. Global invariance is a property of any unitary evolution that is energy conserving and time reversal invariant. Factorisability is a property that quantifies when two parts of a global system can be considered well defined subsystems. The battery states considered in the AQC are parameterized by a map known as the Gibbs map. To derive the AQC we will first introduce these four concepts: the Gibbs map, the quantum time reversal operation, global invariance and factorisability.  

\subsection{Gibbs Map}

The Gibbs map is a mathematical generalisation of the thermal state that turns out to be a convenient means of parameterising the battery states.
\begin{defn} Gibbs map.\\
Given a system with Hamiltonian $H$ at temperature $T$, the action of the Gibbs map, $G_H$, on a state $\rho$ is,
\begin{equation}\label{eq:Gibbsmap}
\begin{aligned}
&G_H(\rho) := \frac{\exp\left(-\frac{H}{2 k_B T}\right) \rho \exp \left(-\frac{H}{2 k_B T}\right)}{\tilde{Z}_H(\rho)} \, , \ \   \mbox{with} \\
&\tilde{Z}_H(\rho) := \Tr \left[\exp\left(-\frac{H}{ k_B T}\right) \rho \right] .
\end{aligned}
\end{equation}
\end{defn}
This map arises naturally as a quantum-mechanical version of the Crooks reversal of a Markov process, and is intimately linked with the Petz recovery map for general quantum states\zo{~\cite{petz,hyukjoon}}. It also has a range of physically natural properties. When the energy of the input state is completely certain, as in an energy eigenstate $\ket{E_k}$, the Gibbs map has no effect and
\begin{equation}\label{Eq: Restricted Thermalisation Classical Limit}
\begin{aligned}
G_H\left(\ket{E_k}\bra{E_k}\right) &= \ket{E_k}\bra{E_k} .
\end{aligned}
\end{equation}
However, when the energy of the input state is completely uncertain, as in a maximally mixed state or in an equal superposition, the state output by the Gibbs map has thermally distributed populations. Additionally, the map is non-dephasing and affects the energy populations in the same way irrespective of the coherent properties the state. In this way, the Gibbs map takes a maximally mixed state, $\rho_\I := \frac{1}{N}\I$, to the thermal state,
\begin{equation}\label{Eq:Partition Function Def}
\begin{aligned}
&G_H(\zo{\rho_\I}) = \gamma(H) := \frac{1}{Z_H} \exp\left(-\frac{H}{k_B T}\right) \, , \ \mbox{with}  \\
&Z_H := \Tr\left[\exp\left(-\frac{H}{k_B T}\right)\right] \, ,
\end{aligned}
\end{equation}
and an equal superposition of energy states, $\ket{\psi_+} := \frac{1}{\sqrt{N}} \sum_{k=1}^N \ket{E_k}$, to the coherent thermal state (a pure state with the same energy populations as the thermal state), 
\begin{equation}
\begin{aligned}
&G_H\left( \ket{\psi_+}\bra{\psi_+} \right) = \zo{\ket{\psi^{\gamma}}  \bra{\psi^{\gamma}}} \, , \ \mbox{with}\\
&\zo{\ket{\psi^{\gamma}}}= \frac{1}{\sqrt{Z_H}}\sum_k^N \exp\left(-\frac{E_k}{2 k_B T}\right)\ket{E_k} \ .
\end{aligned}
\end{equation}
As such we see that the Gibbs maps makes a state crudely `as thermal as possible' subject to the constraints imposed by the input state. However, this loose claim is not intended to be taken literally but rather as a signpost towards the map's deeper physical significance.

It is worth noting that while the Gibbs map is invertible for bounded Hamiltonians; for unbounded Hamiltonians, the normalisation constant $\tilde{Z}_H(\rho)$ is not guaranteed to be finite and so the map is not strictly invertible. However, this does not cause any difficulties in our work because for coherent, squeezed and cat states $\tilde{Z}_H(\rho)$ is finite. A full analysis of the invertibility of the Gibbs map is outside the scope of this paper.

\subsection{Time Reversal}\label{Sec: Time Reversal}

\zrewrite{In broad terms the time reversal operation enacts a motion reversal. This can be visualised as taking a video of the dynamics of a process and running it in reverse. While there are numerous subtleties involved with making this intuitive picture precise~\cite{timerev1, timerev2}; in general terms the classical time reversal operation amounts to the replacement of $t$ by $-t$ in all pertinent physical variables describing the dynamics of a system. 

As time is a parameter rather than observable in quantum mechanics, the quantum time reversal operation acts indirectly on $t$ via its action on the position and momentum operators. To maintain correspondence with classical time reversals, the position operator $x$ is required to remain invariant under a reversal while the momentum operator $p$ changes sign, i.e. $\mathfrak{T}(x) = x$ and $\mathfrak{T}(p) = -p$ \zo{where $\mathfrak{T}(X)$ to denotes the time-reversal of any quantum operator $X$}.
In virtue of the time reversal being a symmetry operation, $\mathfrak{T}$ is additionally required to be length preserving in the sense that \zo{$\Tr[ \ket{\psi} \bra{\psi} ] =  \Tr[\mathfrak{T}(\ket{\psi} \bra{\psi})]$ for any state $\ket{\psi}$}. These two requirements entail that $\mathfrak{T}$ is an anti-unitary operator~\cite{timereversalop}.

The simplest choice in anti-unitary operation to represent $\mathfrak{T}$ is complex conjugation. This is the `textbook'~\cite{timereversalop} quantum time reversal operation; however, the anti-linearity of conjugation can make it arduous to work with. On Hermitian operators, and therefore on any quantum state or observable, the transpose operation is equivalent to complex conjugation. \zo{As the AQC quantifies transition probabilities between quantum states, we are free to use this equivalence and represent $\mathfrak{T}$ via the transpose.}

\zo{\begin{defn} Quantum time reversal.\\
The quantum time reversal operation $\mathfrak{T}$ on any quantum operator $X$ is defined as
\begin{equation}\label{eq:timereversaloperator}
\begin{aligned}
\mathfrak{T} (X) := X^T \, ,
\end{aligned}
\end{equation}
where $^T$ is the transpose operation. It follows that the time reversal operation $\T$ on any pure state $\ket{\psi}$ is given by 
 \begin{equation}\label{eq:timereversepure}
 \T \ket{\psi} =  \ket{\psi^*}  \, ,
 \end{equation}
where $^*$ denotes complex conjugation. 
\end{defn}}

\zo{As the transpose operation is basis dependent, so are $\mathfrak{T}$ and $\T$. The physical situation dictates the appropriate choice in basis. Specifically, the basis is chosen to ensure that the time reversed battery states drive the change in Hamiltonian from $H_S^f$ back to $H_S^i$ and to ensure that the system and battery Hamiltonian, and therefore the system and battery evolution operator, are time reversal invariant. For the osccilator battery and two level system implementation discussed in the main text we take the time reversal operation of the battery with respect to the Fock basis, $\{ \ket{n}\bra{n} \}$, and the time reversal of the system with respect to the Pauli $z$ basis, $\{ \ket{e}\bra{e}, \ket{g}\bra{g} \}$. This flips the sign of the momentum of any battery oscillator state while leaving its position unchanged and ensures that the thermal states of the system are time reversal invariant.}

The derivation of the AQC makes use of two properties of $\mathfrak{T}$. The first of these is the fact that as probabilities are scalars they are invariant under $\mathfrak{T}$. The second is that $\mathfrak{T}$ reverses the order of products of operators, $\mathfrak{T}(A B ) = \mathfrak{T}(B)\mathfrak{T}(A)$.}


\subsection{Global Invariance}\label{Sec: Generalised Detailed Balance}

Global invariance is a property of transition probabilities for a \textit{single} system with a time reversal invariant Hamiltonian and an energy conserving and time reversal invariant unitary evolution. As such, it applies to a large class of systems because most pertinent physical situations are time reversal invariant and all are ultimately unitary and energy conserving when the total setup considered is made sufficiently large. Global invariance is the starting point to derive a large family of quantum fluctuation theorems and plays an analogous role to detailed balance for classical fluctuation theorems.

An energy conserving evolution is here defined to be an evolution that commutes with the Hamiltonian, {\it i.e.} $[U, H]=0$. This is stronger  than requiring the average energy of the setup to remain constant as a result of the evolution. In particular, it rules out the creation of non-degenerate superpositions of energy states from energy eigenstates~\cite{catalyticworkextraction}. 

\change{For the autonomous evolution of a system under a time independent Hamiltonian $H$, the strict energy conservation condition is trivially satisfied as $[\exp(- i H t/\hbar), H] = 0$ and the time reversal invariance of the evolution operator follows from the time reversal invariance of the Hamiltonian, i.e. $\mathfrak{T}(H) = H $ entails $\mathfrak{T}\left(\exp(- i H t/\hbar)\right) = \exp(- i H t/\hbar)$.} 

\medskip 
\medskip


\begin{thm} Global Invariance.\\
Consider a Hamiltonian $H$ and evolution $U$ such that: $\mathfrak{T}(H) = H$, $\mathfrak{T}(U) = U$ and $[U,H] = 0$. Global invariance relates a forwards and reverse transition probability for such a system. In the forwards process, the system is prepared in a state $G_H(\rho_i)$, evolves under $U$ and a binary POVM measurement is performed with POVM elements $\{ \rho_f, \I - \rho_f \}$. The evolved state, $U G_{H}(\rho_i) U^\dagger$, collapses onto $\rho_f$ with the probability
\begin{equation}\label{eq: globaltransfor} 
P(\rho_f | G_{H}(\rho_i)) := \Tr[ \rho_f U G_{H}(\rho_i) U^\dagger] \ .
\end{equation}
In the reverse process the state $G_H(\mathfrak{T}(\rho_f))$ evolves under $U$ and on measurement collapses onto $\mathfrak{T}(\rho_i)$ with the probability
\begin{equation}\label{eq: globaltransrev} 
P(\mathfrak{T}(\rho_i)|G_{H}(\mathfrak{T}(\rho_f))) := \Tr[ \mathfrak{T}(\rho_i) U G_{H}(\mathfrak{T}(\rho_f)) U^\dagger] \ .
\end{equation}
The ratio of these transition probabilities is,
\begin{equation}\label{eq: globalinvariance}
\frac{P(\rho_f | G_{H}(\rho_i))}{P(\mathfrak{T}(\rho_i)|G_{H}(\mathfrak{T}(\rho_f)))} = \frac{\tilde{Z}_H(\mathfrak{T}(\rho_f))}{\tilde{Z}_H(\rho_i)} \ .
\end{equation}
\end{thm}
The proof of global invariance starts with the definition of the forwards transition probability, and makes use of the definition of $\mathfrak{T}$ in combination with the fact that $\mathfrak{T}(H)= H$ and $\mathfrak{T}(U)=U$,
\change{\begin{equation}
\begin{aligned}
P(\rho_f | G_{H}(\rho_i)) &:= \Tr[ \rho_f U G_{H}(\rho_i) U^\dagger] = \Tr[ \mathfrak{T}\left(\rho_f U G_{H}(\rho_i) U^\dagger\right)]  = \Tr[ U^\dagger G_{H}(\mathfrak{T}(\rho_i)) U \mathfrak{T}(\rho_f)] \ .
\end{aligned}
\end{equation}}
If we now substitute in the definition of the Gibbs map, Eq.~\eqref{eq:Gibbsmap}, we have
\begin{equation}
\begin{aligned}
P(\rho_f | G_{H}(\rho_i))  &= \frac{1}{\tilde{Z}_H(\rho_i)} \Tr\left[ U^\dagger\exp\left(-\frac{H}{2 k_B T}\right)\mathfrak{T}(\rho_i)\exp\left(-\frac{H}{2 k_B T}\right) U \mathfrak{T}(\rho_f)\right] \, ,
\end{aligned}
\end{equation}
which can be reordered using the cyclic nature of the trace operation and the fact that $[U, H] =0$,
\begin{equation}
\begin{aligned}
P(\rho_f | G_{H}(\rho_i)) &= \frac{1}{\tilde{Z}_H(\rho_i)} \Tr\left[ \mathfrak{T}(\rho_i)\exp\left(-\frac{H}{2 k_B T}\right) U \mathfrak{T}(\rho_f) U^\dagger \exp\left(-\frac{H}{2 k_B T}\right) \right] \\ &= \frac{1}{\tilde{Z}_H(\rho_i)} \Tr\left[ \mathfrak{T}(\rho_i) U \exp\left(-\frac{H}{2 k_B T}\right)\mathfrak{T}(\rho_f)\exp\left(-\frac{H}{2 k_B T}\right)  U^\dagger  \right] \ .
\end{aligned}
\end{equation}
The proof is completed by reusing the definition of the Gibbs map and the definition of the reverse transition probability,
\begin{equation}
\begin{aligned}
P(\rho_f | G_{H}(\rho_i)) &= \frac{\tilde{Z}_H(\mathfrak{T}(\rho_f))}{\tilde{Z}_H(\rho_i)} \Tr\left[ \mathfrak{T}(\rho_i) U G_H(\mathfrak{T}(\rho_f))  U^\dagger  \right] \\ &= \frac{\tilde{Z}_H(\mathfrak{T}(\rho_f))}{\tilde{Z}_H(\rho_i)} P(\mathfrak{T}(\rho_i)|G_{H}(\mathfrak{T}(\rho_f))) \, .
\end{aligned}
\end{equation}

\zo{The AQC is derived by applying global invariance to a bipartite setup. By evaluating the forwards and reverse transition probabilities,  $P(\rho_f | G_{H}(\rho_i))$ and $P(\mathfrak{T}(\rho_i)|G_{H}(\mathfrak{T}(\rho_f)))$, with respect to a carefully chosen Hamiltonian $H$ and states $\rho_i$ and $\rho_f$, they can be use to quantify a Crooks-like scenario where an initially thermal \textit{system} is driven from equilibrium with a time dependent Hamiltonian with the energy supplied by a quantum \textit{battery}. However, for the initial thermal states of the system to be well defined, the system and battery need to be effectively non-interacting at the start and end of the process. This requirement is captured by an additional assumption that we call the `factorisability' of the system and battery. }


\subsection{Factorisability}\label{Sec: Factorisability}

Factorisability characterises the extent to which multiple, potentially interacting, systems can be considered independent subsystems. It is defined in terms of the map 
\begin{equation}
J_{H}(\sigma) :=\exp\left(-\frac{H}{2 k_B T}\right) \sigma \exp \left(-\frac{H}{2k_B T}\right) 
\end{equation}
for any operator $\sigma$. This map is, on any quantum state $\rho$, the unnormalised version of the Gibbs map, i.e. $J_H( \rho) = \tilde{Z}_H (\rho) \, G_H( \rho) $. 
\begin{defn}Factorisability.\\
A composite system with a Hamiltonian $H_{AB}$ in the state $\rho_A \otimes \rho_B$ is factorisable if
\begin{equation}\label{Eq: factorisability}
\begin{aligned}
& J_{H_{AB}}(\rho_A \otimes \rho_B) = J_{H_A}(\rho_A) \otimes J_{H_B}(\rho_B) 
\end{aligned}
\end{equation}
for a pair of Hamiltonians $H_A$ and $H_B$ of subsystems $A$ and $B$ respectively. 
\end{defn}
\zrewrite{If a composite system is factorisable then the Gibbs map and the Gibbs map normalisation term, Eq.~\eqref{eq:Gibbsmap}, decompose into terms acting on the separate subsystems, i.e. if a composite system with a Hamiltonian $H_{AB}$ in the state $\rho_A \otimes \rho_B$ is factorisable then
\begin{equation}
\begin{aligned}
& G_{H_{AB}}(\rho_A \otimes \rho_B) = G_{H_A}(\rho_A) \otimes G_{H_B}(\rho_B) \ \ \ \ \text{and} \\
& \tilde{Z}_{H_{AB}}(\rho_A \otimes \rho_B) = \tilde{Z}_{H_A}(\rho_A) \otimes \tilde{Z}_{H_B}(\rho_B) \ .
\end{aligned}
\end{equation}

\medskip
To ensure that system and battery are effectively non-interacting at the start and end of non-equilibrium process, we require that given an appropriate choice in battery states their Hamiltonian, $H_{SB}$, factorises into an initial or final system Hamiltonian, $H_S^i$ or $H_S^f$, and battery Hamiltonian, $H_B$. Specifically we assume that the following factorisability condition holds for the system and battery.
\begin{assumpt} Factorisability of the system and battery. \\
There exist a pair of pure battery states, $\ket{\Psi_i}$ and $\ket{\Psi_f}$, such that the system and battery Hamiltonian $H_{SB}$ factorises as
\begin{equation}\label{eq: factoriseassump}
\begin{aligned}
&J_{H_{SB}}\left(\rho_{\I} \otimes \ket{\Psi_i}\bra{\Psi_i} \right) = J_{H_S^i}(\rho_{\I}) \otimes J_{H_B}(\ket{\Psi_i}\bra{\Psi_i}) \ \ \ \  \text{and} \\  
&J_{H_{SB}}\left(\rho_{\I} \otimes \ket{\Psi_f}\bra{\Psi_f} \right) = J_{H_S^f}\left(\rho_{\I}) \otimes J_{H_B}(\ket{\Psi_f}\bra{\Psi_f}\right) \  
\end{aligned}
\end{equation}
where $H_S^i$ and $H_S^f$ are Hamiltonians of the system and $H_B$ is a Hamiltonian of the battery. 
\end{assumpt}
This condition essentially encodes that the system is a genuine system with concrete initial and final Hamiltonians. The AQC follows directly from global invariance, Eq.~\eqref{eq: globalinvariance}, when the factorisability condition, Eq.~\eqref{eq: factoriseassump}, is obeyed. 

\subsection{Main Derivation}

To see how the AQC follows from the factorisability of the system and battery, we set $\ket{\Psi_i}\bra{\Psi_i} = \mathfrak{T}(\ket{\psi_i}\bra{\psi_i} )$ and $\ket{\Psi_f}\bra{\Psi_f} = \mathfrak{T}(\ket{\phi_f}\bra{\phi_f})$ and for convenience define
\begin{equation}\label{eq: statesdef}
    \begin{aligned}
    &\sigma_i = \rho_{\I} \otimes \mathfrak{T}\left(\ket{\psi_i}\bra{\psi_i}\right) \ \ \ \text{and} \\ 
    &\mathfrak{T}(\sigma_f)= \rho_{\I} \otimes \mathfrak{T}(\ket{\phi_f}\bra{\phi_f}) \ .
    \end{aligned}
\end{equation}
The factorisability condition, Eq.~\eqref{eq: factoriseassump}, entails that the Gibbs map acting on $\sigma_i$ and $\mathfrak{T}(\sigma_f)$, Eq. ~\eqref{eq: statesdef}, decomposes as
\begin{equation}
\begin{aligned}\label{eq: factoriseexample}
&G_{H_{SB}}\left(\sigma_i \right) = \gamma_{i}  \otimes \ket{\phi_i}\bra{\phi_i}  \ \ \text{and}  \\
&G_{H_{SB}}\left(\mathfrak{T}(\sigma_f)\right) = \gamma_{f}  \otimes  \ket{\psi_f}\bra{\psi_f} 
\end{aligned}
\end{equation}
where we have used the shorthand $\gamma_k \equiv \gamma(H_S^k)$ for $k = i, f$ and with 
\begin{equation}
    \begin{aligned}\label{eq: relationbetweenstates}
        &\ket{\phi_i}\propto \T  \exp \left(-\frac{H_B}{2 k_B T} \right) \ket{\psi_i} \\
    &\ket{\psi_f}\propto \T  \exp \left(-\frac{H_B}{2 k_B T} \right) \ket{\phi_f} 
    \end{aligned}
\end{equation}
as in the main text. In this way the factorised Gibbs map generates both the thermal states of the system, Eq.~\eqref{eq:thermalstates}, and the temperature dependent operation that parameterises the battery states, Eq.~\eqref{eq:forwardtoreverse}, that, as we will see, appear in the AQC.

We can use the decomposition of the Gibbs map, Eq~\eqref{eq: factoriseexample}, to convert a transition probability of the system \textit{and} battery to a transition probability of the battery alone. The forwards transition probability $P(\sigma_f | G_{H}(\sigma_i))$,  Eq~\eqref{eq: globaltransfor}, of the system and battery reduces to $\mathcal{P}(\phi_f | \phi_i, \gamma_i)$,
the transition probability of the battery from $\ket{\phi_i}$ to $\ket{\phi_f}$ for a system initialised in the thermal state $\gamma_{i}$,
\begin{equation}\label{eq: forreeval}
\begin{aligned}
P\left( \sigma_f  \, | \, G_{H_{SB}}\left( \sigma_i \right)\right) &= \Tr \left[\left(\rho_\I \otimes \ket{\phi_f}\bra{\phi_f}\right) U \,  G_{H_{SB}} \left( \rho_\I \otimes \mathfrak{T}(\ket{\psi_i}\bra{\psi_i}) \right) U^\dagger\right] \\ &= \frac{1}{N} \bra{\phi_f} \Tr_S \left[ U ( \gamma_i \otimes \ket{\phi_i}\bra{\phi_i} ) U^\dagger\right]  \ket{\phi_f} \\ 
&:= \frac{1}{N} \mathcal{P}(\phi_f | \phi_i, \gamma_i) \ .
\end{aligned}
\end{equation}
(The factor $\frac{1}{N}$ is a normalisation term from the maximally mixed system state $\rho_\I$ and cancels out in the final fluctuation theorem.)
Similarly, the reverse global transition probability $ P\left( \mathfrak{T}(\sigma_i) \,|\, G_{H_{SB}}(\mathfrak{T}(\sigma_f)\right)$ reduces to the transition probability of the battery for the reverse process $\mathcal{P}(\psi_i|\psi_f, \gamma_f)$, in which the system is initialised in the thermal state $\gamma_{f}$, 
\begin{equation}\label{eq: revreeval}
    \begin{aligned}
 P\left( \mathfrak{T}(\sigma_i) \,|\, G_{H_{SB}}(\mathfrak{T}(\sigma_f)\right) &= \Tr \left[(\rho_\I \otimes \ket{\psi_i}\bra{\psi_i}) U \, G_{H_{SB}}\left( \rho_\I \otimes \mathfrak{T}(\ket{\phi_f}\bra{\phi_f})\right) U^\dagger \right] \\
&= \frac{1}{N} \bra{\psi_i} \Tr_S \left[U ( \gamma_f \otimes \ket{\psi_f}\bra{\psi_f} ) U^\dagger \right] \ket{\psi_i} \\
&:= \frac{1}{N} \mathcal{P}(\psi_i|\psi_f, \gamma_f) \ .
    \end{aligned}
\end{equation}

It also follows from the factorisation of the system and battery that the Gibbs map normalisation terms decompose as
\begin{equation}
\begin{aligned}\label{eq: factoriseZ}
&\tilde{Z}_{H_{SB}}\left(\sigma_f\right) = \frac{1}{N} Z_{f} \times \tilde{Z}_{\phi_f} \ \ \ \ \text{and} \\
&\tilde{Z}_{H_{SB}}\left(\sigma_i \right) = \frac{1}{N} Z_{i} \times  \tilde{Z}_{\psi_i} \, , 
\end{aligned}
\end{equation}
where we have used the shorthand $Z_k \equiv Z_{H_S^k}$ for $k = i, f$ for the partition functions of the system and $\tilde{Z}_{\phi_f} \equiv \tilde{Z}_{H_B}(\ket{\phi_f}\bra{\phi_f}) $ and $\tilde{Z}_{\psi_i} \equiv \tilde{Z}_{H_B} (\mathfrak{T}(\ket{\psi_i}\bra{\psi_i})) = \tilde{Z}_{H_B} (\ket{\psi_i}\bra{\psi_i})$ as shorthand for the Gibbs map normalisation terms of the battery. 

The AQC now follows directly from substituting the factorised global transition probabilities, Eq.~\eqref{eq: forreeval} and Eq.~\eqref{eq: revreeval}, and the factorised Gibbs map normalisation terms, Eq.~\eqref{eq: factoriseZ}, into global invariance, Eq.~\eqref{eq: globalinvariance}, 
\begin{equation}\label{eq: AQCpartitionfunc}
\begin{aligned}
\frac{\mathcal{P}(\phi_f | \phi_i, \gamma_i)}{\mathcal{P}(\psi_i|\psi_f, \gamma_f)} &= \frac{Z_{f}}{Z_{i}} \frac{\tilde{Z}_{\phi_f} }{\tilde{Z}_{\psi_i}}  \ . 
\end{aligned}
\end{equation}
The derivation is completed by rewriting the ratio of the partition functions in terms of the change in free energy of the system,
\begin{equation}
    \begin{aligned}\label{eq: partitiontofreeenergy}
    \Delta &F := F(H^f_S) - F(H^i_S) \ \ \text{with} \\
    &F(H) : = - k_B T \, \zo{\ln}\left( Z_H  \right) \, ,
    \end{aligned}
\end{equation}
and by analogously defining and the generalised energy flow in terms of the Gibbs map normalisation terms, 
\begin{equation}\label{eq: tildeEandGibbs}
\begin{aligned}
&\Delta \tilde{E} := \tilde{E}(\psi_i) - \tilde{E}(\phi_f)  \ \ \text{with} \\
&\tilde{E}(\psi) : = - k_B T \, \zo{\ln}\left(  \tilde{Z}_{\psi}\right) \,  =  - k_B T \, \zo{\ln}\left(  \bra{\psi} \exp\left(-\frac{H_B}{k_B T}\right) \ket{\psi}\right) .
\end{aligned}
\end{equation}
The AQC can thus be stated as
\begin{equation}\label{eq: AQC}
\begin{aligned}
\frac{\mathcal{P}(\phi_f | \phi_i, \gamma_i)}{\mathcal{P}(\psi_i|\psi_f, \gamma_f) } = \exp \left(-\frac{  \Delta F}{k_B T} \right) \exp \left(\frac{\Delta \tilde{E}}{k_B T} \right)   \ . 
\end{aligned}
\end{equation} }


\zo{\paragraph*{Relationship to Ref. \cite{aberg}}
The above equality, Eq.~\eqref{eq: AQC}, is Eq.(38) of Ref.~\cite{aberg} but written in the limit where factorisability holds exactly and for an initially thermal system and a battery that is prepared in a pure state. Our \textit{battery} plays the role of both the energy reservoir and control system in Ref.~\cite{aberg}.} 

\subsection{\zrewrite{Autonomy and approximate factorisability}}

\zrewrite{To realise an autonomous change in system Hamiltonian we consider a system and a battery with a Hamiltonian of the form
\be \label{eq: explicit hamiltonian}
H_{SB} = H_S^i\otimes \Pi_B^i +  H_B^f\otimes \Pi_B^f + V_{SB}^\perp + \I_S\otimes H_B \ \, ,
\ee
where $\Pi_B^i$ and $\Pi_B^f$ are projectors onto two orthogonal subspaces, $R_i$ and $R_f$, of the battery's Hilbert space, and $V_{SB}^\perp$ has support only outside those two subspaces, {\it i.e.} $(X_S\otimes\Pi_B^i)V_{SB}^\perp=(X_S\otimes\Pi_B^f)V_{SB}^\perp=0$ for any system operator $X_S$. When the battery is initialised in a state in subspace $R_i$ only and evolves to a final state in subspace $R_f$ only, the system Hamiltonian evolves from $H_S^i$ to $H_S^f$. The reverse change in system Hamiltonian $H_S^f$ to $H_S^i$ is similarly realised by a battery initialised in $R_i$ that evolves to subspace $R_f$.

This Hamiltonian, Eq.~\eqref{eq: explicit hamiltonian}, is exactly factorisable on the states $\sigma_i$ and $\mathfrak{\sigma_f}$, Eq.~\eqref{eq: statesdef}, when: (i) the battery state $\T \ket{\psi_i}$ has support solely in $R_i$ and the battery state $\T \ket{\phi_f}$ has support solely in $R_f$, and 
(ii) the battery Hamiltonian does not induce evolution between subregions, i.e. $ (\I_B - \Pi_B^i ) H_B \Pi_B^i = 0$ and $ (\I_B - \Pi_B^f ) H_B \Pi_B^f = 0$. Under these two restrictions it is straightforward to verify that
\begin{equation}\label{eq: factorise-i}
\begin{aligned}
J_{H_{SB}}(\sigma_i) &= \exp\left(-\frac{H_S^i\otimes \Pi_B^i + \I_S \otimes H_B}{2 k_B T}\right) (\rho_{\I} \otimes \mathfrak{T}( \ket{\phi_f}\bra{\phi_f} )) \exp\left(-\frac{H_S^i\otimes \Pi_B^i + \I_S \otimes H_B}{2 k_B T}\right) \\ 
&= J_{H_S^i}(\rho_{\I}) \otimes J_{H_B}(\mathfrak{T}(\ket{\phi_f}\bra{\phi_f})) \ \ \ \ \ \  
\end{aligned}
\end{equation}
and similarly for $J_{H_{SB}}(\mathfrak{T}(\sigma_f))$. Thus in this limit the AQC holds exactly.  

\medskip

However, to realise an \textit{autonomous} change in system Hamiltonian, the evolution of the battery between subspaces $R_i$ and $R_f$ must be generated by the unitary evolution of the system and battery under $H_{SB}$, \textit{i.e.} by the propagator $U = \exp(- i H_{SB} t/ \hbar) $. Consequently, in practise $H_B$ must induce evolution between the subregions $R_i$ and $R_f$ and so $ (\I_B - \Pi_B^i ) H_B \Pi_B^i \neq 0$ and $ (\I_B - \Pi_B^f ) H_B \Pi_B^f \neq 0$. As a result, the system and battery factorisability condition, Eq.~\eqref{eq: factoriseassump}, does not hold exactly for autonomous dynamics and consequently nor does the AQC. 

The AQC is thus strictly only ever approximate. Its approximate nature stems from the fact that when considering the autonomous dynamics of a system and battery under a time-independent interacting Hamiltonian, the system never strictly has a well-defined local Hamiltonian. This tension is implicit to all fluctuation theorems and arises from the impossibility of perfectly delineating the individual energies of interacting systems. Nonetheless, the tension is rarely a practical problem because it is generally possible to identify approximately well defined subsystems with their own concrete Hamiltonians.

The AQC holds to a high degree of approximation whenever it is possible to identify the system and battery as approximately well defined subsystems with approximately well defined Hamiltonians at the start and end of the forwards and reverse protocols. This idea is captured by the claim that with an appropriate choice in the battery states $\ket{\psi_i}$ and $\ket{\phi_f}$ it is possible to ensure that factorisability holds to a high degree of approximation, \textit{i.e.}
\begin{equation}\label{eq: factoriseassumpapprox}
\begin{aligned}
J_{H_{SB}}(\sigma_i) &\simeq J_{H_S^i}(\rho_{\I}) \otimes J_{H_B}( \mathfrak{T}(\ket{\psi_i}\bra{\psi_i})) \ \ \ \  \text{and} \\ 
J_{H_{SB}}(\mathfrak{T}(\sigma_f)) &\simeq J_{H_S^f}(\rho_{\I}) \otimes J_{H_B}(\mathfrak{T}(\ket{\phi_f}\bra{\phi_f}))  \ .
\end{aligned}
\end{equation}
From which it follows that with an appropriate choice in the battery states $\ket{\psi_i}$ and $\ket{\phi_f}$ the AQC also holds to a high degree of approximation
\begin{equation}\label{eq: AQCapprox}
\begin{aligned}
\frac{\mathcal{P}(\phi_f | \phi_i, \gamma_i)}{\mathcal{P}(\psi_i|\psi_f, \gamma_f) } \simeq \exp \left(-\frac{  \Delta F}{k_B T} \right) \exp \left(\frac{\Delta \tilde{E}}{k_B T} \right)   \ . 
\end{aligned}
\end{equation}

\medskip

The error-bounded AQC, Eq.~\eqref{eq: Error Bounded Equality}, is an inequality that is obeyed when even when the factorisability condition does not hold exactly. The extent to which factorisability holds can be quantified by the difference between left and right hand side of the system and battery factorisability condition, Eq.~\eqref{eq: factoriseassumpapprox}. We thus define the initial and final factorisation errors,  $\epsilon_{SB}^{i}$ and $\epsilon_{SB}^{f}$, as
\begin{equation}\label{Eq: error bound}
\begin{aligned}
&\epsilon_{SB}^{i} := J_{H_{SB}}(\I_S \otimes \ket{\psi_i}\bra{\psi_i}) -   J_{H_{S}^{i}}(\I_S) \otimes J_{H_B}(\ket{\psi_i}\bra{\psi_i})  \ \  \ \text{and} \\ &\epsilon_{SB}^{f}:= J_{H_{SB}}(\I_S \otimes \ket{\phi_f}\bra{\phi_f})  -   J_{H_S^f}(\I_S) \otimes J_{H_B}(\ket{\phi_f}\bra{\phi_f}).
\end{aligned}
\end{equation}
The proof of the error-bounded AQC proceeds in the same manner as the proof of the AQC except that we keep track of the error generated from factorisability holding imperfectly. We start with the global invariance condition for the states $\sigma_i$ and $\sigma_f$, Eq.~\eqref{eq: statesdef},
\begin{equation}
\begin{aligned}
\tilde{Z}_{H_{SB}}(\rho_\I \otimes \ket{\psi_i}\bra{\psi_i}) \, P(\rho_\I \otimes \ket{\phi_f}\bra{\phi_f} | G_{H_{SB}}(\rho_\I \otimes \mathfrak{T}(\ket{\psi_i}\bra{\psi_i}))) \\ = \tilde{Z}_{H_{SB}}(\rho_\I \otimes \ket{\phi_f}\bra{\phi_f}) \, P(\rho_\I \otimes \ket{\psi_i}\bra{\psi_i}|G_{H_{SB}}(\rho_\I \otimes \mathfrak{T}(\ket{\phi_f}\bra{\phi_f}))) \, .
\end{aligned}
\end{equation}
and substitute in the definition of the factorisability errors $\epsilon_{SB}^{i}$ and $\epsilon_{SB}^{f}$, Eq. \ref{Eq: error bound},
\begin{equation}
\begin{aligned}
&Z_{i} \, \tilde{Z}_{\psi_i} \,  \mathcal{P}(\phi_f | \phi_i, \gamma_i) + \Tr[(\I_S \otimes \ket{\phi_f}\bra{\phi_f}) U \mathfrak{T}(\epsilon_{SB}^i) U^\dagger ] \\ &=  Z_{f} \, \tilde{Z}_{\phi_f} \, \mathcal{P}(\psi_i|\psi_f, \gamma_f) + \Tr[(\I_S \otimes \ket{\psi_i}\bra{\psi_i}) U \mathfrak{T}( \epsilon_{SB}^f ) U^\dagger ]  \, ,
\end{aligned}
\end{equation}
which on rearranging and using the triangle inequality gives
\begin{equation}
\begin{aligned}
& |Z_i \, \tilde{Z}_{\psi_i} \, \mathcal{P}(\phi_f | \phi_i, \gamma_i) - Z_f \,  \tilde{Z}_{\phi_f} \, \mathcal{P}(\psi_i|\psi_f, \gamma_f) | \\ & \le | \Tr[(\I_S \otimes \ket{\phi_f}\bra{\phi_f}) U \mathfrak{T}(\epsilon_{SB}^i) U^\dagger ] | + | \Tr[(\I_S \otimes \ket{\psi_i}\bra{\psi_i}) U \mathfrak{T}(\epsilon_{SB}^f) U^\dagger ] | \ .
\end{aligned}
\end{equation}
It follows from the Cauchy Schwartz identity that
\begin{equation}
\begin{aligned}
& \Tr[ ( U^\dagger (\I_S \otimes \ket{\phi_f}\bra{\phi_f}) U ) \, \epsilon_{SB}^i ] \leq  ||\mathfrak{T}(\epsilon_{SB}^i) ||_\zo{1}  = ||\epsilon_{SB}^i ||_\zo{1}\ \ \ \text{and} \\
& \Tr[ ( U^\dagger (\I_S \otimes \ket{\psi_i}\bra{\psi_i}) U ) \, \epsilon_{SB}^f ] \leq  ||\mathfrak{T}(\epsilon_{SB}^f )||_\zo{1} =||\epsilon_{SB}^f ||_\zo{1}  \, . 
\end{aligned}
\end{equation}
Thus we are left with the error-bounded AQC,
\begin{equation}\label{Eq: Error Bounded Autonomous Crooks}
\begin{aligned}
| Z_i \, \tilde{Z}_{\psi_i} \, \mathcal{P}(\phi_f | \phi_i, \gamma_i) - Z_f \, \tilde{Z}_{\phi_f} \, \mathcal{P}(\psi_i|\psi_f, \gamma_f) | \le ||\epsilon_{SB}^f||_\zo{1}+|| \epsilon_{SB}^i||_\zo{1} := \epsilon \ .
\end{aligned}
\end{equation}
  }

\paragraph*{Non-autonomous variant.}
It is possible to obtain an equality of the same general form as the AQC, Eq.~\eqref{eq: AQC}, for a non-autonomous setup where the evolution is induced by an \textit{applied} unitary $U$ rather than by evolution under $H_{SB}$, i.e. $U \neq \exp(- i H_{SB} t/\hbar )$. This is in fact the central result of Ref.~\cite{aberg}. In the non-autonomous variant an additional control system $C$ is introduced and the total Hamiltonian (up to technicalities concerning time reversals) is of the form,
\be \label{eq: new total hamiltonian}
H_{CSB}  = \ket{C_i}\bra{C_i} \otimes H_S^i\otimes \I_B  +   \ket{C_f}\bra{C_f} \otimes H_S^f\otimes \I_B+ V_{CSB}^\perp\ + \I_C \otimes \I_S\otimes H_B\, 
\ee
where these control states are orthogonal, $\braket{C_f |C_i} = 0$, and $ V_{CSB}^\perp$ has orthogonal support to both $\ket{C_i}$ and $\ket{C_f}$, i.e. $ V_{CSB}^\perp\ (\ket{C_i}\bra{C_i} \otimes X_S \otimes X_B) = V_{CSB}^\perp\ (\ket{C_f}\bra{C_f} \otimes X_S \otimes X_B) =0$ for any system and battery operators $X_S$ and $X_B$. A change in system Hamiltonian $H_S^i$ to $H_S^f$ is induced by applying a unitary that evolves the control from the state $\ket{C_i}$ to $\ket{C_f}$. By requiring the applied unitary to be energy conserving and time reversal invariant, global invariance holds as before. Furthermore, as the above Hamiltonian, Eq.~\eqref{eq: new total hamiltonian}, is effectively non-interacting as long as the control is prepared in the state $\ket{C_i}$, or the state $\ket{C_f}$, the factorisability condition holds exactly. Given that these two conditions are satisfied the rest of the derivation runs identically to that of the AQC. The resulting quantum Crooks equality holds \textit{exactly} and differs from the AQC only by the particular definition of the transition probabilities. It follows that the coherent, squeezed and cat state equalities can also be used for this non-autonomous setup. 

\medskip

\paragraph*{The role of the thermal bath.}

In this paper we have treated the thermal bath as implicit: a bath is a means of preparing the system in the thermal state but we do not model it. Effectively this amounts to assuming that the system and bath are sufficiently weakly interacting at the start of the protocols that they can be considered well defined independent systems at these times. 

More rigorously, we can think of the system discussed in this paper as an enlarged system consisting of some small system of interest (s) that is driven by a change in Hamiltonian and a thermal bath (b) with a constant Hamiltonian $H_b$, i.e. $H_{S}^i = H_s^i \otimes H_b$ and  $H_{S}^f = H_s^f \otimes H_b \,$. The AQC can be re-derived explicitly including the thermal bath if an additional factorisability condition is assumed to hold between the system and bath. The factorisability condition will hold if their shared Hamiltonian is non-interacting in the regions $R_i$ and $R_f$, i.e.
\begin{equation}
H_{sbB} = H_{s}^i \otimes H_{b} \otimes \Pi_B^i +  H_{s}^f \otimes H_{b} \otimes \Pi_B^f + V_{sbB}^\perp\ + \I_{sb}\otimes H_B \ .
\end{equation}
Under these circumstances, the influence of the bath `factorises out'. The resulting equality explicitly quantifies the transition probabilities of the battery, when a system is driven with a change in Hamiltonian, in the presence of a thermal bath. This quantum Crooks equality takes exactly the same form as the usual AQC, Eq.~\eqref{eq:qcrooks}, but with the relevant transition probabilities replaced by
\begin{equation}
\begin{aligned}
&\mathcal{P}(\phi_f | \phi_i, \gamma_i):= \bra{\phi_f} \Tr_{sb} \left[ U_{sbB} ( \gamma(H_s^f) \otimes \gamma(H_b) \otimes \ket{\phi_i}\bra{\phi_i} ) U_{sbB}^\dagger\right]  \ket{\phi_f} \ \ \ \text{and} \\
&\mathcal{P}(\psi_i|\psi_f, \gamma_f) := \bra{\psi_i} \Tr_{sb} \left[U_{sbB} ( \gamma(H_s^f) \otimes \gamma(H_b) \otimes \ket{\psi_f}\bra{\psi_f} ) U_{sbB}^\dagger \right] \ket{\psi_i}  \ .
\end{aligned}
\end{equation}

A worthwhile extension to our research would be to explicitly simulate the bath in the experimental proposal and numerical analysis. One means of doing so would be to investigate the master equation fluctuation theorem that is proposed in the final version of~\cite{aberg}. Alternatively one could try and incorporate the bath using the quantum jump approach explored in~\cite{autonomousfluc}.

\subsection{Properties of generalised energy flow}\label{Appendix: properties of tilde}

Fundamentally, the term $\tilde{E}$ appears in the AQC because the energy of a general quantum state is not well defined and instead some statistical estimate of the states' energy is required. The precise form of $\tilde{E}$ is forced by the fluctuation theorem approach of comparing a forwards and reverse process, i.e. by the structure of the derivation. In a contrast to Eq.~\eqref{Eq: tildeE} in the main text, we here treat the Hamiltonian $H$ and temperature $T$ as variables,
\be\label{Eq: tildeEexplicitHT}
 \tilde{E}(\zo{\psi}, H, T) := -k_B T \, \zo{\ln}\left( \zo{\bra{\psi}} \exp\left(- \frac{ H}{k_B T}\right) \zo{\ket{\psi}} \right)\, ,
\ee
to enable us to discuss the properties of the function $\tilde{E}$ more generally. An analysis of $\tilde{E}$ provides some physical support for our interpretation of $\Delta \tilde{E}$ as a quantum generalisation of the energy flow between the system and the battery. In particular, the following properties hold:

\begin{enumerate}
\item $ \tilde{E}(\psi, H + \delta,  T) = \tilde{E}(\psi, H, T) + \delta $.
\item  $ \tilde{E}(\psi, \lambda H, T) = \lambda \tilde{E}(\psi, H, \lambda T)  $.
\item For an energy eigenstate, $\ket{E_k}$, $\tilde{E}$ is simply the associated eigenstate energy, $\tilde{E} (E_k, H, T) = E_k $. 
\item  In the high temperature limit $\tilde{E}$ tends to the average energy of the state, $ \lim_{T\to\infty} \tilde{E}(\psi, H, T) = \bra{\psi} H \ket{\psi}$.
\item In the general case $\tilde{E}$ is less than the average energy: $ \tilde{E}(\psi, H, T) \leq \bra{\psi} H \ket{\psi}$.
\item $\tilde{E}$ is independent of the the phase of the state $ \tilde{E}(\psi, H ,T ) = \tilde{E}\left(\psi e^{i \phi}, H, T \right) \ \forall \ \phi \in \Re$.
\end{enumerate}

Properties 1 and 2 verify that $\tilde{E}$ scales as one would expect an energy measure to scale. Property 3 is intuitive because when the energy of a state is well-defined there is no need to statistically estimate it. Moreover, property 3 is required to regain the classical limit. Property 6 ensures that in the absence of interactions, $\tilde{E}$ is constant in time which is a desirable condition for a statistical estimate of the energy of a quantum state.

\section{Derivation of the coherent, squeezed and cat state AQCs}\label{Appendix: Corollary Algebra}

\subsection{Coherent state AQC}\label{Appendix: Coherent States}
The coherent state AQC is derived by considering a harmonic oscillator battery, i.e. $H_B =  \hbar \omega \left( a^\dagger a + \frac{1}{2} \right) = \sum_n \hbar \omega \left( n + \frac{1}{2} \right) \ket{n} \bra{n}$, and transition probabilities between coherent states. The measured states $\ket{\phi_f}$ and $\ket{\psi_i}$ are set to the coherent states $\ket{\alpha_{f}}$ and $\ket{\alpha_{i}^*}$ respectively,
\begin{equation}\label{Eq: coherent equality derivation0}
\begin{aligned}
&\ket{\phi_f} = \ket{\alpha_{f}} \ \ \text{and} \\
&\ket{\psi_i} = \ket{\alpha_{i}^*} \, ,
\end{aligned}
\end{equation}
where $\ket{\alpha} := \exp(-\frac{}{}|\alpha|^2/2) \sum_{n=0}^{\infty} \frac{\alpha^n}{n!} \, \ket{n}$.
We calculate $\ket{\phi_i}$ and $\ket{\psi_f}$ using the relation between states $\ket{\phi_{i,f}}$ and $\ket{\psi_{i,f}}$ as specified in Eq.~\eqref{eq:forwardtoreverse}. This requires calculating the effect of applying the temperature dependent operation $\exp\left(-\frac{H_B}{2 k_B T}\right) = \exp\left(-\chi \left(a^\dagger a + \frac{1}{2}\right)\right) $ where $\chi = \frac{\hbar \omega}{2 k_B T}$, to a coherent state. Using operator algebra we find that
\begin{equation}\label{Eq: coherent equality derivation1}
\begin{aligned}
& \ket{\phi_i} = \frac{1}{\sqrt{\tilde{Z}_{\alpha_i}}} \exp(-|\alpha_i|^2/2) \sum_n \exp\left(-\chi \left(a^\dagger a + \frac{1}{2}\right)\right) \frac{\alpha_i^n}{n!} \ket{n} = \ket{\alpha_i \exp\left(-\chi \right)} \ \ \text{and} \\ 
&\ket{\psi_f} = \T \left(\frac{1}{\sqrt{\tilde{Z}_{\alpha_f}}}  \exp(-|\alpha_f|^2/2) \sum_n \exp\left(-\chi \left(a^\dagger a + \frac{1}{2}\right)\right) \frac{\alpha_f^n}{n!} \ket{n} \right) = \ket{\alpha_f^* \exp\left(-\chi \right)} \ . 
\end{aligned}
\end{equation}
Additionally, given the definition of $\Delta \tilde{E}$ in Eq.~\eqref{eq:flow}, it can be shown using operator algebra that 
\begin{equation}\label{Eq: coherent equality derivation2}
\begin{aligned}
\exp\left(\frac{\Delta \tilde{E}}{k_B T}\right) =  \frac{\tilde{Z}_{\alpha_f}}{\tilde{Z}_{\alpha_i}}  = \frac{\left(\frac{\exp\left(-|\alpha_f|^2/2\right)}{\exp(-|\alpha_f \exp(- \chi)|^2/2)}\right)^{2}}{ \left(\frac{\exp(-|\alpha_i|^2/2)}{\exp\left(-|\alpha_i \exp\left(-\chi \right)|^2/2\right)}\right)^2  } =  \exp \left((|\alpha_f|^2 -|\alpha_i|^2)(\exp(- 2\chi)-1) \right) \ . 
\end{aligned}
\end{equation}

The error-bounded coherent state AQC follows from Eq.~\eqref{Eq: coherent equality derivation0}, Eq.~\eqref{Eq: coherent equality derivation1} and Eq.~\eqref{Eq: coherent equality derivation2} and the error bounded AQC, Eq.~\eqref{eq: Error Bounded Equality}.

\zo{To compare the coherent state Crooks equality to the classical Crooks equality, Eq. \ref{eq:Crooks}, we rewrite $\Delta \tilde{E}$ in terms of the difference between the average energy change of the \textit{battery} in the forwards and reverse processes, $W_q:=(\Delta E_+ - \Delta E_-)/2$, with
    \begin{equation}
    \begin{aligned}
         &\Delta E_+ := \bra{\phi_i} H_B \ket{\phi_i} - \bra{\phi_f} H_B \ket{\phi_f} \ \  \  \text{and}  \\
         &\Delta E_- := \bra{\psi_f} H_B \ket{\psi_f} - \bra{\psi_i} H_B \ket{\psi_i} \,  .
    \end{aligned}
    \end{equation}
    Since energy is globally conserved, $-W_q$ is also the difference between the average energy changes of the \textit{system} in the forwards and reverse processes. When the battery is prepared and measured in energy eigenstates, $W_q$ reduces from an average to a well defined energy change. 
    Specifically, if $\ket{\psi_{i,f}} = \ket{E_{i,f}}$ then it follows from Eq.~\eqref{eq: relationbetweenstates} that $ \ket{\phi_{i,f}} = \ket{E_{i,f}}$ and therefore 
    \begin{equation}
    W_q = \frac{1}{2}\left( ( E_i - E_f ) - -( E_i - E_f )\right) \equiv \frac{1}{2} \left( W- (-W)\right) = W 
    \end{equation}
    where $(E_i - E_f) \equiv W$ is the energy gained (lost) by the system (battery).
    Beyond this classical limit, $W_q$ is a generalisation of $W$ that accounts for coherence by considering the battery's average energy and allows for the fact that the AQC quantifies different average energy gains/losses in the forwards and reverse processes (i.e. that in general $\Delta E_+ \neq -\Delta E_-$).}

By calculating $\Delta E_+$ and $\Delta E_-$ for coherent states,
\begin{align}
\Delta E_+ := (\exp(-2\chi) |\alpha_i|^2 - |\alpha_f|^2) \hbar \omega \\
\Delta E_- := (\exp(-2\chi)|\alpha_f|^2 - |\alpha_i|^2) \hbar \omega \, ,
\end{align}
we can rewrite $\Delta \tilde{E}$ for coherent states, Eq.~\ref{Eq: coherent equality derivation2}, in terms of $W_q$, 
\begin{equation}
\begin{aligned}
\Delta \tilde{E} = k_B T (|\alpha_i|^2 -|\alpha_f|^2)(1 -  \exp(-2\chi)) = \frac{1}{\chi}\frac{1 - \exp(-2 \chi) }{1+\exp(-2\chi)} \frac{1}{2}(\Delta E_+ - \Delta E_- ) = \frac{1}{\chi}\frac{1 - \exp(-2 \chi) }{1+\exp(-2\chi)} W_q  \ .
\end{aligned}
\end{equation}
Thus the coherent state AQC can be written as
\begin{equation}
\begin{aligned}\label{eq: coherentstaterewriteqfactor}
&\frac{\mathcal{P}\left( \alpha_f |  \alpha_i \exp(-\chi), \gamma_i \right)}{\mathcal{P}\left(\alpha_i^* |  \alpha_f^* \exp(-\chi), \gamma_f \right)}   
= \exp \left( - \frac{ \Delta F }{k_B T}\right) \, \exp \left(q\left(\chi\right)\frac{W_q}{k_B T}\right) \, ,
\end{aligned}
\end{equation} 
where we have defined the quantum prefactor,
\begin{equation}\label{eq: prefactor}
q\left(\chi \right) := \frac{1}{\chi}\frac{1 - \exp(-2\chi) }{1+\exp(-2\chi)} = \frac{1}{\chi} \tanh(\chi) \ .
\end{equation}

The quantum prefactor, Eq.~\eqref{eq: prefactor}, can be related to the average energy of a thermal battery $\langle H_B \rangle_{\gamma(H_B)} :=\Tr[ H_B \gamma(H_B)] $. We calculate the explicit algebraic form of $\langle H_B \rangle_{\gamma(H_B)}$ and find that
\begin{equation}
\begin{aligned}
\langle H_B \rangle_{\gamma({H_B})} &= \sum_n \hbar \omega (n+1/2) \frac{\exp\left(-\frac{\hbar \omega(n + \frac{1}{2})}{k_B T}\right)}{Z_{H_B}}\\&= \frac{1}{2}\hbar \omega \left( \frac{1+\exp(-2\chi)}{1-\exp(-2\chi)}\right)
\\&= \frac{k_B T}{q(\chi)} \ .
\end{aligned}
\end{equation}
Rearranging we are left with
\begin{equation}\label{eq: qfactor redefined}
q\left(\chi \right) = \frac{k_B T}{\langle H \rangle_{\gamma(H_B)}} := \frac{k_B T}{\hbar \omega_{T}} \ .
\end{equation}
where we have implicitly defined the `thermal frequency' $\omega_{T}$ as the average frequency of a harmonic oscillator in a thermal state. The rewritten coherent state Crooks equality in the main text, Eq.~\eqref{eq: coherentstaterewrite}, follows from Eq.~\eqref{eq: coherentstaterewriteqfactor} and Eq.~\eqref{eq: qfactor redefined}.


\medskip 
\paragraph*{Comment on classical limit.} For the coherent state Crooks equality, the classical limit corresponds to $\chi = 
\frac{\hbar \omega}{2 k_B T} \rightarrow 0$. In this limit, the core physics of the coherent state Crooks equality is equivalent to that of the classical Crooks equality; however, there are technical distinctions resulting from their different formulations. In particular, the coherent state Crooks equality quantifies state transitions of the battery rather than energy changes of the system and these are related many-to-one. However, in the classical limit the coherent state equality quantifies transitions between battery states with sharp energies and as energy is globally conserved this corresponds to the sharp energy changes of the system quantified by the classical Crooks equality. As such, battery states can be viewed as pointer states to determine the energy flow in or out of the system during the protocols.

\subsection{Cat state Crooks equality}\label{Appendix: Cat States}
We take a cat state to be an equal superposition of two arbitrary coherent states, $\ket{\mbox{\small Cat}} \propto ( \ket{\alpha} + \ket{\beta})$. This is a more general definition than that sometimes seen in the literature where a cat state is taken to be only a superposition of a pair of coherent states $\ket{\alpha}$ and $\ket{-\alpha}$. To derive a cat state AQC the measured states $\ket{\phi_f}$ and $\ket{\psi_i}$ are set to
\begin{equation}
\begin{aligned}
\ket{\phi_f} &= \ket{\mbox{\small Cat}_f} := \frac{1}{\sqrt{2+ 2\Re(\exp(-\frac{1}{2}(|\alpha_f|^2 + |\beta_f|^2 -2 \beta_f^* \alpha_f)))}}  (\ket{\alpha_f} + \ket{\beta_f} ) \ \ \text{and} \\
\ket{\psi_i} &= \ket{\mbox{\small Cat}_i^*} :=\frac{1}{\sqrt{2+ 2\Re(\exp(-\frac{1}{2}(|\alpha_i|^2 + |\beta_i|^2 -2 \beta_i^* \alpha_i)))}} (\ket{\alpha_i^*} + \ket{\beta_i^*} ) \, ,
\end{aligned}
\end{equation}
where the normalisation term is calculated using the fact that the overlap between two coherent states $\ket{\alpha}$ and $\ket{\beta}$ is given by $\braket{\beta | \alpha} = \exp(-\frac{1}{2}(|\alpha|^2 + |\beta|^2 -2 \beta^* \alpha))$.
The prepared states $\ket{\phi_i}$ and $\ket{\psi_f}$ are calculated using the relation between states $\ket{\phi_{i,f}}$ and $\ket{\psi_{i,f}}$ as specified in Eq.~\eqref{eq:forwardtoreverse}. Using operator algebra we find that
\begin{equation}\label{eq: CatGibbs}
\begin{aligned}
\ket{\phi_i} &:=   \exp\left(-\chi \left(a^\dagger a + \frac{1}{2}\right)\right) \ket{\mbox{\small Cat}_i} \propto \left(\exp \left(-\chi a^\dagger a\right) \ket{\alpha_i} + \exp \left(-\chi a^\dagger a\right)  \ket{\beta_i} \right) \\
&= \frac{1}{\sqrt{\tilde{Z}_{\mbox{\tiny Cat}_i}}} \left( \eta_\chi^{|\alpha_i|^2} \ket{\exp(-\chi) \alpha_i} + \eta_\chi^{|\beta_i|^2}  \ket{\exp(-\chi) \beta_i} \right)  \ \ \text{and} \\
\ket{\psi_f} &= \frac{1}{\sqrt{\tilde{Z}_{\mbox{\tiny Cat}_f}}} \left(\eta_\chi^{|\alpha_f|^2} \ket{\exp(-\chi) \alpha_f^*} + \eta_\chi^{|\beta_f|^2} \ket{\exp(-\chi) \beta_f^*} \right)
\end{aligned}
\end{equation}
where we have defined
\begin{equation}
\begin{aligned}
\eta_\chi &:= \exp\left(-\frac{1}{2}\left(1- \exp(-2\chi)\right)\right) \ 
\end{aligned}
\end{equation}
and the normalisation term $\tilde{Z}_{\mbox{\tiny Cat}_i}$ (and equivalently for $\tilde{Z}_{\mbox{\tiny Cat}_f}$) takes the form
\begin{equation}\label{eq: catZtilde}
\begin{aligned}
\tilde{Z}_{\mbox{\tiny Cat}_i} = \frac{ \eta_\chi^{2|\alpha_i|^2} 
+ \eta_\chi^{2|\beta_i|^2} + \eta_\chi^{|\alpha_i|^2+|\beta_i|^2}2\Re\left(\exp\left(-\frac{1}{2}\exp(-2 \chi)(|\beta_i|^2 + |\alpha_i|^2 -2 \beta_i^* \alpha_i) \right)\right)}{2+ 2\Re(\exp(-\frac{1}{2}(|\alpha_i|^2 + |\beta_i|^2 -2 \beta_i^* \alpha_i)))} \  .
\end{aligned}
\end{equation}
The normalisation terms $\tilde{Z}_{\mbox{\tiny Cat}_i}$ and $\tilde{Z}_{\mbox{\tiny Cat}_f}$ also provide the $\exp\left( \frac{\Delta \tilde{E} }{k_B T}\right)$ term. It follows from the definition of $\Delta \tilde{E}$ in Eq.~\eqref{eq:flow} that
\begin{equation}
\begin{aligned}
&\exp\left( \frac{\Delta \tilde{E} }{k_B T} \right) = \frac{\bra{\phi_f} \exp\left(-2 \chi \right)\ket{\phi_f}}{\bra{\psi_i}\exp\left(- 2 \chi \right)\ket{\psi_i}} = \frac{\tilde{Z}_{\mbox{\tiny Cat}_f}}{\tilde{Z}_{\mbox{\tiny Cat}_i}}  \ .
\end{aligned}
\end{equation}
As such, the cat state AQC follows from Eq.(B12 -B16). (We do not state the explicit form of $\exp\left( \frac{\Delta \tilde{E} }{k_B T} \right) $ in terms of $\alpha_{i,f}$, $\beta_{i,f}$, $\chi$ and $\eta_\chi$ as the expression does not simplify.) 
\medskip
The error-bounded cat state Crooks equality follows from the error bounded autonomous Crooks equality, Eq.~\eqref{Eq: Error Bounded Autonomous Crooks}, with $\tilde{Z}_{i,f}$ defined in Eq.~\eqref{eq: catZtilde}. 


\subsection{Squeezed state Crooks equality}\label{Appendix: Squeezed States}

The squeezed state AQC quantifies transition probabilities between battery oscillator states that are not only displaced but also squeezed. As such, the measured states $\ket{\phi_f}$ and $\ket{\psi_i}$ are set to the squeezed displaced states $\ket{\alpha_{f}, r_f} $ and $\ket{\alpha_{i}^*, r_i}$ respectively, i.e.
\begin{equation}
\begin{aligned}
& \ket{\phi_f}= \ket{\alpha_{f}, r_f} = D(\alpha_{f})S(r_{f})\ket{0} \ \ \text{and} \\
& \ket{\psi_i} =\ket{\alpha_{i}^*, r_i} = D(\alpha_{i}^*)S(r_{i})\ket{0} \, ,
\end{aligned}
\end{equation}
where we have introduced the displacement operator $D(\alpha) = \exp(\alpha a^\dagger - \alpha^* a)$,  the squeeze operator $S(r) = \exp(\frac{r}{2}(a^2 - {a^\dagger}^2))$, and for simplicity we assume from the outset that the squeezing parameters $r_i$ and $r_f$ are real. We calculate $\ket{\phi_i}$ and $\ket{\psi_f}$ using the relation between states $\ket{\phi_{i,f}}$ and $\ket{\psi_{i,f}}$ as specified in Eq.~\eqref{eq:forwardtoreverse}. This amounts to calculating the effect of applying the operator $\exp \left(-\chi a^\dagger a\right)$ to a squeezed displaced state, i.e.
\begin{equation}\label{Eq: squeeze displace calc}
\begin{aligned}
& \ket{\phi_i} = \frac{1}{\sqrt{\tilde{Z}_{\alpha_i, r_i}}}   \exp\left(-\chi \left(a^\dagger a + \frac{1}{2}\right)\right) D(\alpha_{i})S(r_{i})\ket{0} \equiv D(\mu_{i})S(t_{i})\ket{0} = \ket{\mu_{i}, t_{i}} \ \ \text{and} \\
& \ket{{\psi_f}^*} = \frac{1}{\sqrt{\tilde{Z}_{\alpha_f, r_f}}}  \exp\left(-\chi \left(a^\dagger a + \frac{1}{2}\right)\right) D(\alpha_{f}^*)S(r_{f})\ket{0} \equiv D(\mu_{f}^*)S(t_{f})\ket{0} = \ket{\mu_{f}^*, t_{f}} \ .
\end{aligned}
\end{equation}
We introduce $\mu_{i,f}$ and $t_{i,f}$ to denote the displacement and squeezing parameters respectively of the prepared states. (That the states $\ket{\phi_i}$ and $\ket{\psi_{f}^*} $ are also squeezed displaced states is a non-trivial result of the calculation.) 
To derive the AQC we will also need to calculate the $\exp \left(\frac{\Delta \tilde{E}}{k_B T} \right)$ term. It follows from the definition of $\Delta \tilde{E}$ in Eq.~\eqref{eq:flow} that this can be written as
\begin{equation}
\begin{aligned}
\exp \left(\frac{\Delta \tilde{E}}{k_B T}\right) = \frac{\tilde{Z}_{\alpha_f, r_f}}{\tilde{Z}_{\alpha_i, r_i}} = \frac{(\exp(-\chi a^\dagger a)  D(\alpha_{f})S(r_{f})\ket{0})^\dagger \exp(-\chi a^\dagger a)  D(\alpha_{f})S(r_{f})\ket{0}}{(\exp(-\chi a^\dagger a)  D(\alpha_{i}^*)S(r_{i})\ket{0})^\dagger \exp(-\chi a^\dagger a)  D(\alpha_{i}^*)S(r_{i})\ket{0}} \ .
\end{aligned}
\end{equation}
To derive the AQC we thus need to calculate $\mu$, $s$ and $\tilde{Z}_{\alpha, r}$.

\subsubsection{Useful Identities}

To calculate $\mu$, $s$ and $\N_{\alpha, r}$ we will make use of the following equalities:

\small

\begin{enumerate}
\item $\exp(m a^\dagger a ) \exp(n a^\dagger) \exp(-m a^\dagger a)  = \exp(n \exp(m) a^\dagger)$ 
\begin{proof}
This is derived using the Taylor expansion of the exponential followed by the Hadamard lemma~\cite{identityref},
\begin{equation*}
\begin{aligned}
\exp(m a^\dagger a) \exp(n a^\dagger) \exp(-m a^\dagger a)  =  \sum_k \frac{(n \exp(m a^\dagger a ) a^\dagger \exp(-m a^\dagger a) )^k}{k!} = \sum_k \frac{(n a^\dagger \exp(m) )^k}{k!}  =  \exp(n \exp(m) a^\dagger) \ .
\end{aligned} 
\end{equation*}
\end{proof}

\item $\exp(m a ) \exp( n a^\dagger) = \exp(mn) \exp( n a^\dagger) \exp(m a )$
\begin{proof}
This is derived using the Baker-Campbell-Hausdorff formula~\cite{identityref} followed by the Zassenhaus formula~\cite{identityref},
\begin{equation*}
\exp(m a ) \exp( n a^\dagger) = \exp(m a + n a^\dagger) \exp\left(\frac{m n}{2}\right) =  \exp( n a^\dagger) \exp(m a ) \exp(mn) \ . 
\end{equation*}
\end{proof}

\item $ \exp(m a^2) \exp(n a^\dagger) = \exp(m n^2) \exp(n a ^\dagger) \exp(m a^2) \exp(2 m n a) $
\begin{proof}
This is again derived using the Baker-Campbell-Hausdorff formula followed by the Zassenhaus formula,
\begin{equation*}
\begin{aligned}
 \exp(m a^2) \exp(n a^\dagger) = \exp\left(\frac{m n^2}{6}\right)\exp(m a^2 + mn a + n a^\dagger) =  \exp(m n^2) \exp(n a^\dagger) \exp(m a^2) \exp(2mn a) 
\end{aligned}
\end{equation*}
\end{proof}

\item $\exp \left(\frac{1}{2}(ma^2 + n{a^\dagger}^2 )\right) = \frac{1}{\sqrt{\cos \left(\sqrt{m n}\right)}} \exp\left(\frac{1}{2}\sqrt{\frac{n}{m}}\tan(\sqrt{mn}){a^\dagger}^2\right) \exp\left(- \zo{\ln}\left(\cos(\sqrt{mn})\right)a^\dagger a\right) \exp\left(\frac{1}{2} \sqrt{\frac{m}{n}}\tan(\sqrt{mn}) a^2\right)$ 
\begin{proof}
To derive this equality it is helpful to first introduce the operators $K_+ := \frac{1}{2}{a^\dagger}^2$, $K_-:=\frac{1}{2}a^2$, $K_z := \frac{1}{4}(a a^\dagger + a^\dagger a)$. These operators obey the commutation relations $[K_+, K_-]= -2K_z$ and $[K_z, K_\pm] = \pm K_\pm$. Next, we introduce the ansatz 
\begin{equation}\label{eq: ansatz}
\begin{aligned}
f := \exp((m K_+ + n K_-)t) = \exp(p K_+) \exp(q K_z) \exp(s K_-) \ .
\end{aligned}
\end{equation}
To find the coefficients $p$, $q$ and $s$ we differentiate $f$ with respect to $t$,
\begin{equation}\label{eq: differentiated}
\begin{aligned}
f' &= (m K_+ + n K_-) f = \left( p' K_+ + q' \exp(p K_+)  K_z \exp(-p K_+) + s' \exp(pK_+)\exp(q K_z) K_- \exp(-q K_z)  \exp(-pK_+) \right) f.
\end{aligned}
\end{equation}
To simplify this expression we use the Hadamard lemma to obtain the following relations,
\begin{equation}\label{eq: hadamard relations}
\begin{aligned}
&\exp(p K_+)  K_z \exp(-p K_+) = K_z - p K_+  \ \ \ \text{and} \\
&\exp(pK_+)\exp(q K_z) K_- \exp(-q K_z) \exp(-pK_+) =  \exp(pK_+) \exp(-q) K_-  \exp(-pK_+) = \exp(-q) (K_- - 2pK_z + p^2 K_+ ) \ . 
\end{aligned}
\end{equation}
Eq.~\eqref{eq: hadamard relations} can be substituted into the right hand side of Eq.~\eqref{eq: differentiated}, from which it then follows that
\begin{equation}\label{eq: equation coeff}
\begin{aligned}
m K_+ + n K_- =  p' K_+ + q' (K_z + p K_+ ) + s' \exp(-q) (K_- - 2pK_z + p^2 K_+ ) \ .
\end{aligned}
\end{equation}
Equating coefficients in Eq.~\eqref{eq: equation coeff} we obtain a set of differential equations,
\begin{equation*}
\begin{aligned}
&p' - q'p + s' \exp(-q) p^2 = m \\
&q' - 2ps' \exp(-q) = 0 \\
&s' \exp(-q) = n \, ,
\end{aligned}
\end{equation*}
which can be solved, subject to the constraint that $p(0) = q(0) = s(0) = 0$, to find that
\begin{equation}\label{eq: differential equations}
\begin{aligned}
&p = \frac{m}{n} \tan \left(\sqrt{m n} t \right)\\
&q = -2 \zo{\ln} \left( \cos \left(\sqrt{mn} t \right)\right) \\ 
&s = \frac{n}{m} \tan \left(\sqrt{m n} t \right) \ .
\end{aligned}
\end{equation}
Identity 4 is obtained by substituting this solution, Eq.~\eqref{eq: differential equations}, back into the ansatz, Eq.~\eqref{eq: ansatz}.
\end{proof}

\item $\exp \left(\frac{1}{2} \left( m a^2 + n {a^\dagger}^2 \right)\right) =\sqrt{\cos(\sqrt{m n})} \exp \left(\frac{1}{2} \sqrt{\frac{m}{n}}\tan(\sqrt{mn}) a^2 \right) \exp \left(\zo{\ln} \left(\cos(\sqrt{mn})\right)a^\dagger a \right) \exp \left(\frac{1}{2}\sqrt{\frac{n}{m}}\tan(\sqrt{mn}){a^\dagger}^2 \right) $

\begin{proof}
The proof here follows the same method as for identity 4. Using the ansatz $f = \exp((m K_+ + n K_-)t) = \exp(p K_-) \exp(q K_z) \exp(s K_+)$ it is possible to derive the same differential equations as in Eq.~\eqref{eq: differential equations} but with $q \rightarrow -q$.
\end{proof}

\item $\exp \left(m a^2 \right) \exp \left(n{a^\dagger}^2 \right) = \frac{1}{\sqrt{1-4mn}} \exp \left(\frac{n}{1-4m n } {a^\dagger}^2 \right) \exp \left(\zo{\ln} \left(\frac{1}{1-4mn} a^\dagger a \right) \right) \exp \left(\frac{m}{1-4mn} a^2 \right)$

\begin{proof}
This is proven using identities 4 and 5. Let $p = \frac{1}{2} \sqrt{\frac{m}{n}}\tan(\sqrt{mn}) $, $q = \frac{1}{2}\sqrt{\frac{n}{m}}\tan(\sqrt{mn})$ and $\exp(- 2 M) = \cos(\sqrt{mn})^2 = \frac{1}{1 - 4 p q}$, it follows that
\begin{equation*}
\sqrt{M} \exp(p a^2)\exp(\zo{\ln}(M) a^\dagger a) \exp(q {a^\dagger}^2) = \frac{1}{\sqrt{M}} \exp(q {a^\dagger}^2 ) \exp(-\zo{\ln}(M) a^\dagger a) \exp(p a^2)  \, ,
\end{equation*}
which can be rearranged to give
\begin{equation*}
\begin{aligned}
\exp \left(-p a^2 \right) \exp \left(q {a^\dagger}^2 \right)  &= M \exp(\zo{\ln}(M a^\dagger a) \exp(q {a^\dagger}^2) \exp(-p a^2)\exp(\zo{\ln}(M) a^\dagger a)  \\ 
&= M  \exp(q \exp(- 2 M) {a^\dagger}^2)  \exp(2\zo{\ln}(M a^\dagger a)  \exp(-p \exp(- 2 M) a^2) \ . 
\end{aligned}
\end{equation*}
The substitution $p \leftrightarrow -p$ then gives identity 6. 
\end{proof}
\end{enumerate} 

\normalsize

\subsubsection{Main Derivation}

We are now in a position to calculate the effect of the operator $\exp \left(- \chi a^\dagger a \right)$ on a squeezed displaced state,
\begin{equation}\label{Eq: start of calc one}
\begin{aligned}
\exp \left(- \chi a^\dagger a \right)D(\alpha) S(r)\ket{0} &= \exp \left(- \chi a^\dagger a\right) \exp(\alpha a^\dagger - \alpha^* a ) \exp \left(\frac{r}{2} \left(a^2 - {a^\dagger}^2 \right) \right) \ket{0} \ .
\end{aligned}
\end{equation}
We first use the standard factorised form of the displacement and squeeze operators~\cite{quantumopticstextbook}, 
\begin{equation}\label{eq: standard factorised form}
D(\alpha) S(r)\ket{0} = \frac{\exp \left(-\frac{1}{2}|\alpha|^2\right)}{\sqrt{\cosh(r)}} \exp\left(\alpha a^\dagger\right)\exp \left(-\alpha^* a\right) \exp\left(-\frac{1}{2}\tanh(r) {a^\dagger}^2\right) \ket{0} 
\end{equation}
to rewrite Eq.~\eqref{Eq: start of calc one},
\begin{equation}\label{Eq: start of calc two}
\begin{aligned}
&= \exp \left(- \chi a^\dagger a \right) \frac{\exp \left(-\frac{1}{2}|\alpha|^2\right)}{\sqrt{\cosh(r)}} \exp\left(\alpha a^\dagger\right)\exp \left(-\alpha^* a\right) \exp\left(-\frac{1}{2}\tanh(r) {a^\dagger}^2\right) \ket{0} \ . 
\end{aligned}
\end{equation}
Eq.~\eqref{Eq: start of calc two} is then rewritten using identity 3,
\begin{equation}
\begin{aligned}
= \frac{\exp\left(-\frac{1}{2}|\alpha|^2\right)}{\sqrt{\cosh(r)}}  \exp \left(-\chi a^\dagger a \right) \exp\left(\alpha a^\dagger\right) \exp\left(-\frac{1}{2} \tanh(r) {\alpha^*}^2\right) \exp \left(\alpha^* \tanh(r) a^\dagger \right) \exp \left(-\frac{1}{2} \tanh(r) {a^\dagger}^2 \right) \ket{0} \, ,
\end{aligned}
\end{equation}
followed by identity 1,
\begin{equation}\label{new parameters}
\begin{aligned}
= \frac{\exp(-\frac{1}{2}(|\alpha|^2-{\alpha^*}^2 \tanh(r)))}{\sqrt{\cosh(r)}} \exp\left( \exp\left(- \chi \right)(\alpha + \alpha^* \tanh(r)) a^\dagger\right) \exp\left(-\frac{1}{2}  \exp\left(- 2 \chi \right)\tanh(r) {a^\dagger}^2\right) \ket{0} \ . 
\end{aligned}
\end{equation}
By comparing Eq.~\eqref{new parameters} with itself in the limit that $\chi = 0$, i.e. the limit in which $\exp \left(- \chi a^\dagger a \right)$ is not applied and Eq.~\eqref{new parameters} is simply a rewritten version of a squeezed displaced state, we identify the following modified displacement and squeeze coefficients $\mu$ and $s$,
\begin{equation}\label{Eq: modified squeeze parameters}
\begin{aligned}
&\Re(\mu) = \Re(\alpha) \frac{\exp\left(- \chi\right)(1 + \tanh(r))}{1 + \exp\left(-2 \chi \right)\tanh(r)} \, , \\
&\Im(\mu) = \Im(\alpha) \frac{\exp\left(- \chi\right)(1 - \tanh(r)}{1 - \exp\left(-2 \chi \right)\tanh(r))} \ \ \text{and} \\
&\tanh(s) = \exp\left(- 2 \chi \right) \tanh(r) \ . 
\end{aligned}
\end{equation}

It remains to calculate the $\exp \left(\frac{\Delta \tilde{E}}{k_B T}\right)$ term. This amounts to calculating,
\begin{equation}
\begin{aligned}
\tilde{Z}_{\alpha, r} &= \left(\exp\left(-\chi a^\dagger a\right)  D(\alpha)S(r)\ket{0}\right)^\dagger \exp \left(-\chi a^\dagger a\right)  D(\alpha)S(r)\ket{0}
\end{aligned}
\end{equation}
The above equation can be immediately simplified using Eq.~\eqref{new parameters}
\begin{equation}
\begin{aligned}
&= \frac{\exp \left(-(|\alpha|^2-\Re(\alpha)^2 \tanh(r))\right)}{\cosh(r)} \bra{0} \exp\left(m {a}^2\right) \exp\left( n a\right)  \exp\left( n^* a^\dagger\right) \exp\left(m^* {{a}^\dagger}^2 \right) \ket{0} 
\end{aligned}
\end{equation}
where
\begin{equation}
\begin{aligned}
&m = -\frac{1}{2} \exp \left(- 2 \chi \right) \tanh(r) \ \text{and} \\
&n = \exp\left(- \chi \right)(\alpha^* + \alpha \tanh(r)) \ .
\end{aligned}
\end{equation}
Using `Useful Identities' 6, 3, 1, and 2 in succession this can be rewritten to give,
\begin{equation}
\begin{aligned}
&\bra{0} \exp\left(m {a}^2\right) \exp\left( n a\right)  \exp\left( m^* a^\dagger\right) \exp\left(n^* a^\dagger \right) \ket{0} \\ &= \frac{1}{r} \bra{0} \exp\left( n a\right) \exp \left(\frac{m^*}{\exp(- 2r)}{a^\dagger}^2 \right) \exp \left(-\zo{\ln} \left(\exp(- 2r)\right)a^\dagger a \right) \exp \left(\frac{m}{\exp(- 2M)} a^2 \right) \exp\left(n^* a^\dagger \right) \ket{0} \\
&= \frac{1}{M} \exp \left(\frac{2\Re(n^2 m)}{\exp(- 2M)}\right)  \bra{0} \exp\left( n a\right) \exp(-ln(\exp(- 2M))a^\dagger a) \exp\left(n^* a^\dagger \right) \ket{0} \\
&= \frac{1}{M} \exp \left(\frac{2\Re(n^2 m)}{\exp(- 2M)}\right)  \bra{0} \exp\left( \frac{n}{M} a\right) \exp\left(\frac{n^*}{M} a^\dagger \right) \ket{0} \\
&= \frac{1}{M} \exp \left(\frac{2\Re(n^2 m)}{\exp(- 2M)}\right) \exp\left(\frac{|n|^2}{\exp(- 2M)}\right) 
\end{aligned}
\end{equation}
where $M = \frac{1}{\sqrt{1 - 4|m|^2}}$. It follows that,
\small \begin{equation}
\begin{aligned}
&\tilde{Z}_{\alpha, r} 
:= \frac{\exp(-(|\alpha|^2-\Re(\alpha^2) \tanh(r)))}{\cosh(r) \sqrt{1 - \tanh(r)^2\exp\left(- 4 \chi \right)}} \\ &\exp \left(\frac{{\exp \left(-2 \chi \right) (|\alpha|^2 (1+ \tanh(r)^2) + 2\tanh(r)\Re(\alpha^2))-\exp\left(-2 \chi \right)\tanh(r)\left(\Re({\alpha}^2)+ \Re(\alpha^2)\tanh(r)^2+ 2|\alpha|^2 \tanh(r)\right)}}{1 - \tanh(r)^2\exp\left(- 4 \chi \right)} \right) \, ,
\end{aligned}
\end{equation}
\normalsize
which can be simplified if we assume that the displacement parameters are real,  $\alpha_{i,f} = \Re(\alpha_{i,f})$,
\begin{equation}
\begin{aligned}
&\tilde{Z}_{\alpha, r} 
:= \frac{\exp \left(-|\alpha|^2(1-\tanh(r) \right)}{\cosh(r) \sqrt{1 - \tanh(r)^2\exp\left(- 4 \chi \right)}} &\exp \left( |\alpha|^2 \frac{(1+\tanh(r))^2\left(1-\tanh(r)\exp\left(- 2 \chi \right)\right)\exp\left(- 2 \chi \right)}{1 - \tanh(r)^2 \exp\left(- 4 \chi \right)} \right) \ .
\end{aligned}
\end{equation}
This completes the derivation of the squeezed state AQC. The error-bounded squeezed state Crooks equality follows from the error bounded AQC, Eq.~\eqref{eq: Error Bounded Equality}.

\section{Dynamics of proposal}
Analysis of the dynamics of the harmonic oscillator state under the proposed interaction, Eq.~\eqref{eq: level splitting interaction}, verifies that the oscillator behaves as a battery. 
The interaction is diagonal in the system energy eigenbasis and thus the total Hamiltonian can be written in the form
\begin{equation}
H_{SB} = \ket{e}\bra{e} \otimes H_B^e  +  \ket{g} \bra{g} \otimes H_B^g \, ,
\end{equation}
where $H_B^e = H_B + E(x_B)$ and $H_B^g =  H_B - E(x_B)$.
Consequently, $H_{SB}$ does not induce transitions between the system energy levels. 
At the start of the forwards protocol the two-level system and oscillator are prepared in the state
\begin{equation}
\rho_{SB}(0) = (p_e \ket{e}\bra{e} + p_g \ket{g}\bra{g} ) \otimes  \ket{\phi_i}\bra{\phi_i} \, ,
\end{equation}
where $\frac{p_e}{p_g} = \exp \left(\frac{-2 E_i} {k_BT}\right)$. The system and battery then evolve under $H_{SB}$, for some time $t$, to the state
\begin{equation}
\begin{aligned}
\rho_{SB}(t) = p_e \ket{e}\bra{e} \otimes \exp(-i t H_B^{e}/\hbar) \ket{\phi_i}\bra{\phi_i} \exp(i t H_B^{e}/\hbar) + p_g \ket{g}\bra{g} \otimes \exp(-i t H_B^{g}/\hbar) \ket{\phi_i}\bra{\phi_i} \exp(i t H_B^{g}/\hbar) .
\end{aligned}
\end{equation}
As such the oscillator state evolves into two non-equally weighted components, $\ket{\phi_{i}^e(t)}  := \exp(-i t H_B^{e}/\hbar) \ket{\phi_i} $ and $ \ket{\phi_{i}^g(t)} := \exp(-i t H_B^{g}/\hbar) \ket{\phi_i} $, correlated with the system being prepared in the excited and ground state respectively, i.e. 
\begin{equation}
\begin{aligned}
\rho_{SB}(t) = p_e \ket{e}\bra{e} \otimes  \ket{\phi_{B}^e(t)}\bra{\phi_{B}^e(t)}  + p_g \ket{g}\bra{g} \otimes \ket{\phi_{B}^g(t)}\bra{\phi_{B}^g(t)} \ .
\end{aligned}
\end{equation}

The oscillator state in the forwards protocol is prepared such that its average position is initially in the region of narrow splitting, i.e. $ \bra{\phi_i} x_B \ket{\phi_i} < x_i $. In this case $\ket{\phi_{B}^e(t)}\bra{\phi_{B}^e(t)}$ travels up a potential hill and by energy conservation its average energy decreases by an amount $E_f - E_i$ as it travels from $x_i$ to $x_f$. Conversely, $\ket{\phi_{B}^g(t)}\bra{\phi_{B}^g(t)}$ travels down a potential hill and its average energy increases by an amount $E_f - E_i$. 

\begin{figure}
\subfloat[Initial State]{\includegraphics[width = 3.4in]{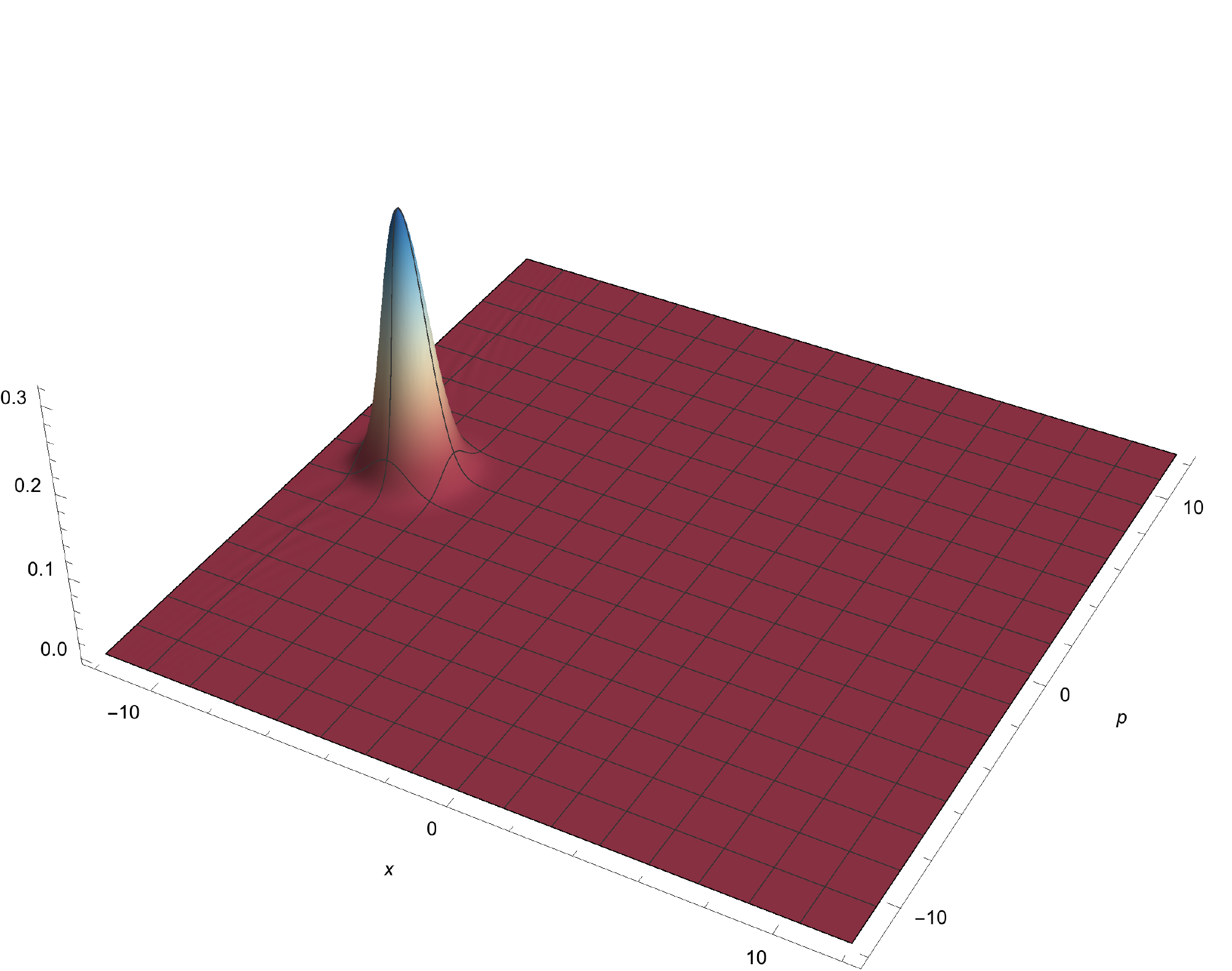}} \hspace{3mm}
\subfloat[Evolved State]{\includegraphics[width = 3.4in]{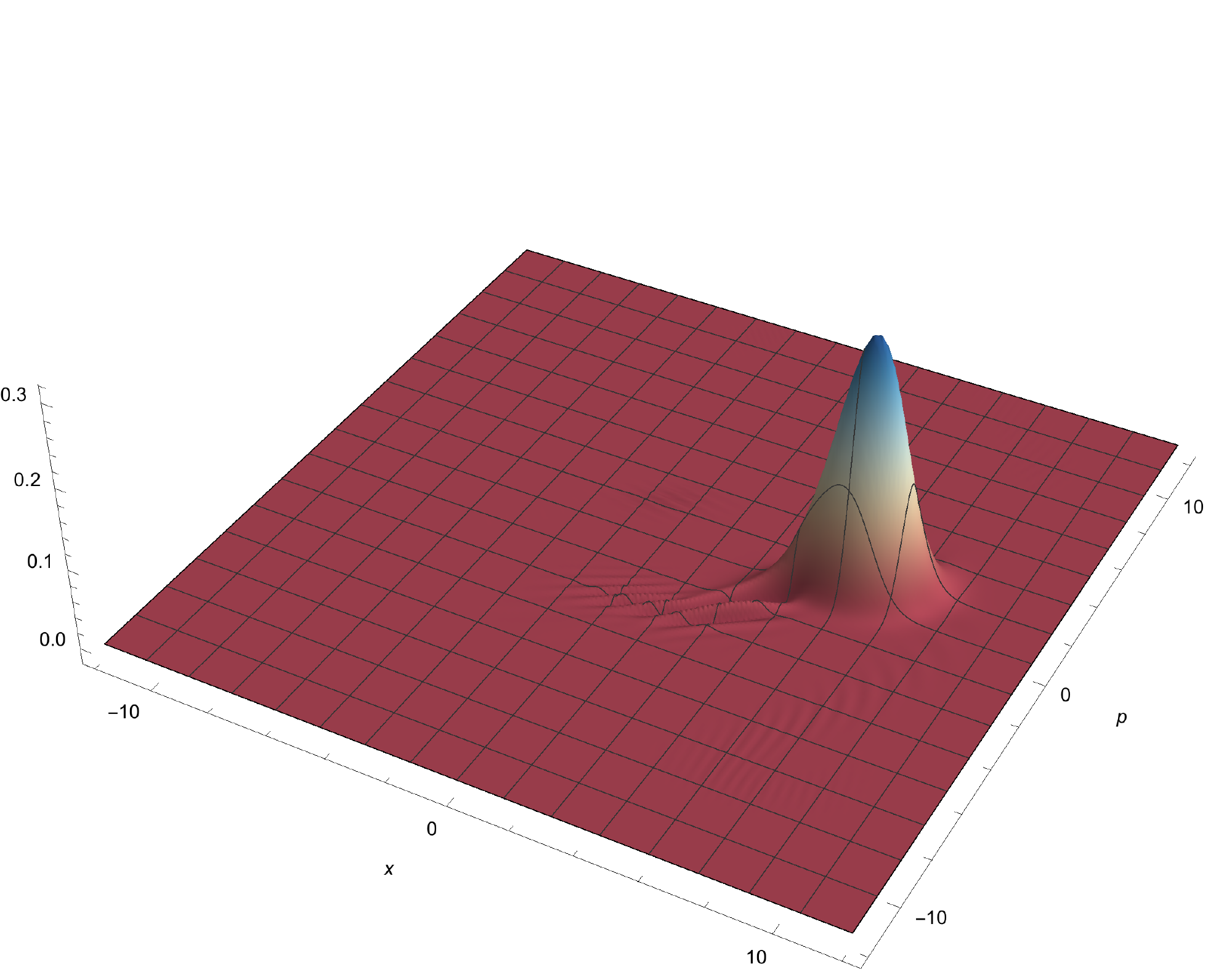}}
\caption{These are plots of the Wigner function of the a) initial oscillator state and b) final oscillator state after evolution under our proposed Hamiltonian. In these plots the following parameters are chosen somewhat arbitrarily: $m=1$, $\hbar \omega = 1$, $E_i= 1$, $E_f = 21$ and the interaction region extends from $x_i = -2$ to $x_f =2$. The oscillator starts in a coherent state centered around $x = -9$. These plots correspond to the case in which the system is prepared in the excited state and so show the component of the wavepacket that travels up the potential hill: $\ket{\phi_{i}^e(t)} := \exp(-i H_B^e t) \ket{\exp(-\chi) \alpha}$.  The oscillator wavepacket is slowed as it evolves through the potential hill and the interaction squeezes the coherent state and leave small residual ripples trailing behind the bulk of the Wigner function. The Wigner function is negative in the troughs of these ripples.}
\label{wigner plots}
\end{figure}

Numerical simulations reveal that the oscillator state wavepacket is distorted as it evolves through the interaction region. In Fig. \ref{wigner plots} we see that a coherent oscillator states is squeezed and interference `ripples' are generated by the interaction. The Wigner function of these `ripples' is in places negative. Negative values of a quasi probability distribution are hard to explain classically so this is a signature of quantum mechanical phenomena. Furthermore, numerical simulations indicate that when the final state is approximated by a coherent state with the same average position and momentum as the complete final wavepacket the error-bounded coherent state AQC does not always hold. As such, we see that even when the battery is prepared in a coherent state, the most classical of the motional quantum states, it does not remain in this approximately classical state. Moreover, the quantum distortion effects are essential to the coherent state AQC being obeyed. 

\end{document}